\documentclass[11pt]{article}
\usepackage[pdftex]{color}
\usepackage{graphicx}
  \usepackage{amsthm,amsfonts}
  \usepackage{amsmath}
\usepackage{slashed}

\usepackage{caption}
\usepackage{fullpage}

\newcommand{\bea}   {\begin{eqnarray}}
\newcommand{\eea}   {\end{eqnarray}}

\begin{document}
\renewcommand{\thefootnote}{\fnsymbol{footnote}}

\thispagestyle{empty}

\title{A three-dimensional superconformal quantum \\mechanics with $sl(2|1)$ dynamical symmetry}
\author{Ivan E. Cunha\thanks{{E-mail: {\em ivanec@cbpf.br}}}\quad and\quad
Francesco
Toppan\thanks{{E-mail: {\em toppan@cbpf.br}}}
\\
\\
}
\maketitle

\centerline{
{\it CBPF, Rua Dr. Xavier Sigaud 150, Urca,}}\centerline{\it{
cep 22290-180, Rio de Janeiro (RJ), Brazil.}}

~\\
\maketitle
\begin{abstract} 

We construct a three-dimensional superconformal quantum mechanics (and its associated de Alfaro-Fubini-Furlan deformed oscillator) possessing an $sl(2|1)$ dynamical symmetry. At a coupling parameter $\beta\neq 0$ the Hamiltonian contains a $\frac{1}{r^2}$ potential and a spin-orbit (hence, a first-order differential operator) interacting term. At $\beta=0$ four copies of undeformed three-dimensional oscillators are recovered. The Hamiltonian
gets diagonalized in each sector of total $j$ and orbital $l$ angular momentum (the spin of the system is $\frac{1}{2}$). The Hilbert space of the deformed oscillator is given by a direct sum of $sl(2|1)$ lowest weight representations. The selection of the admissible Hilbert spaces at given values of the coupling constant $\beta$ is discussed. The spectrum of the model is computed. The vacuum energy (as a function of $\beta$) consists of a recursive zigzag pattern. The degeneracy of the energy eigenvalues grows linearly up to $E\sim \beta$ (in proper units) and quadratically for $E>\beta$. The orthonormal energy eigenstates are expressed in terms of the associated Laguerre polynomials and the spin spherical harmonics. The dimensional reduction of the model to $d=2$ produces two copies (for $\beta$ and $-\beta$, respectively) of the
two-dimensional $sl(2|1)$ deformed oscillator. The dimensional reduction to $d=1$ produces the one-dimensional $D(2,1;\alpha)$ deformed oscillator, with $\alpha$ determined by $\beta$.
~\\\end{abstract}
\vfill
\rightline{CBPF-NF-002/19}

\newpage

\section{Introduction}

In this paper we construct a three-dimensional superconformal quantum mechanics having, as an input, an $sl(2|1)$ dynamical symmetry. The associated  (denoted as ``DFF", see \cite{dff}) de Alfaro-Fubini-Furlan type of deformed oscillator, with $sl(2|1)$ as spectrum-generating superalgebra, is presented in detail. These quantum models (both the superconformal and the DFF one) are defined in terms of a dimensionless deformation parameter $\beta$ which, without loss of generality, can be assumed to belong to the $\beta\geq 0$  interval. The $\beta=0$ undeformed DFF Hamiltonian corresponds to four copies of the three-dimensional isotropic oscillator; when $\beta> 0$, a spin-orbit term enters the Hamiltonians. It is a first-order differential operator which can be diagonalized in each sector of given total and orbital angular momenta.\par
The list of the main results is the following:\\
{\em i}) once derived the $sl(2|1)$ lowest weight representations, alternative admissible Hilbert spaces that can be associated with the quantum models at a given $\beta$ are constructed. Consistency conditions require the wave functions to be normalized and the Hamiltonian to be self-adjoint. The procedure is an extension of the approach and results discussed in \cite{{mt},{ftf}} for conformal quantum mechanics.  The spectrum of the deformed oscillators is computed. When varying the deformation parameter $\beta$ (which, in physical applications, can play the role of an external
control parameter) certain lowest weight representations of $sl(2|1)$ can be ``switched on" as admissible in the
given Hilbert space, while certain other lowest weight representations can be ``switched off". One of the consequences is the production, for the vacuum energies, of the recursive zigzag patterns observed in Figures {\bf 1} and {\bf 2};\\
{\em ii}) an unexpected feature concerning the degeneracy of  the energy eigenvalues of the deformed oscillator is derived. For $\beta$ integer or half-integer the energy spectrum is a shifted version of the spectrum of the undeformed
oscillator.  The $\beta$-deformed oscillator realizes an interpolation between two different regimes.
Up to energy $E\sim\beta$ (measured in natural units, by setting  $\Delta(E)=1$, where $\Delta(E)=E_1-E_{vac}$ is the energy difference
between the first excited state and the vacuum) the degeneracy grows linearly, mimicking the behaviour of a two-dimensional oscillator; starting from $E\sim \beta$, the degeneracy grows quadratically (we recall that the degeneracy $d(n)$ of the $n$-th excited energy eigenvalue of an ordinary  $D$-dimensional oscillator  grows as
$d(n)\propto n^{D-1}$). This behaviour has been computed in Section {\bf 6} and visually presented in Figure {\bf 3}; \\
{\em iii}) the orthonormal eigenstates are expressed in terms of two classical functions: the associated Laguerre polynomials dependent on the (square of) the radial coordinate and the spin spherical harmonics, see \cite{bieden}, dependent on the angular coordinates. This result has been derived in Appendix {\bf B}, see formulas (\ref{eveneigen}) and (\ref{oddeigen}).\par 
Besides the above points, further discussed features are the implementation of superselection rules, the recovery
of lower-dimensional deformed oscillators via dimensional reduction and so on.\par
Conformal quantum mechanics was first introduced in \cite{cal}. The superconformal extension was presented
in \cite{fr} and \cite{ap}. Superconformal quantum mechanics and its associated DFF deformed oscillators are a very active field of investigation with a growing body of literature. There are two main motivations for that. On one side, the development of sophisticated mathematical tools (e.g., for large ${\cal N}$-extended supersymmetry, the role of superconformal Lie algebras) to construct and explicitly solve models, both at the classical and quantum level. On a physical side, for its important applications. We mention in particular the motion of test particles in the proximity of the horizon of certain black holes, see \cite{bp} and the $AdS_{2}/CFT_1$ correspondence investigated in \cite{sen} and \cite{cj}. The $AdS_2$ holography has recently gained new attention in relation with the Sachdev-Ye-Kitaev models (see \cite{ms, kit} and references therein); in this context the connection between conformal and schwarzian mechanics is discussed, e.g., in \cite{msy,sch2}.  \par
In the literature there are three main approaches to construct superconformal quantum mechanics. The most popular one consists in quantizing classical world-line superconformal sigma-models defined on supermultiplets 
of an ${\cal N}$-extended one-dimensional supersymmetry (for ${\cal N}=1,2,4,8$ the relevant supermultiplets,
see \cite{{pato},{kuroto}}, are of the type $(D,{\cal N},{\cal N}-D)$, corresponding to $D$ propagating bosonic fields,
${\cal N}$ fermionic fields and ${\cal N}-D$ auxiliary bosonic fields). In the sigma-model interpretation, $D$ is the dimensionality of the target manifold. These classical superconformal world-line sigma-models are constructed either via superspace (see the \cite{fil} review and the references therein; more recent works are \cite{{ivsi},{fediva},{ils},{ils2}}) or via $D$-module representations of one-dimensional superconformal algebras, as in \cite{{kuto},{khto}}.  \par 
The so-called ``triangular representations" (in contraposition to the ordinary ``parabolic representations") of superconformal algebras have been discussed in \cite{hoto}. They induce the de Alfaro-Fubini-Furlan deformed oscillators counterparts of the superconformal mechanics. Some recent works on superconformal quantum mechanics, either  ``parabolic" or
``trigonometric" cases, are \cite{{su21is},{cuhoto},{sqmfis}, {akt},{su21fils},{ackt}}. Particularly relevant for our purposes here is
the \cite{cuhoto} paper. There, it is shown that the quantization of world-line superconformal sigma-models with $D\geq 3$
target dimensions cannot be performed straightforwardly, but require solving non-trivial non-linear equations. It is due to this obstruction that we apply here a more direct method (the second approach, pionereed in \cite{{akt},{ackt}} for one-dimensional models) to construct a three-dimensional superconformal quantum mechanics. It is rewarding that the dimensional reductions of the three-dimensional superconformal quantum mechanics allow to recover (see Section {\bf 7})  the models obtained in \cite{cuhoto} by quantizing the worldline superconformal sigma models based on the ${\cal N}=4$ $(1,4,3)$ and the ${\cal N}=2$ $(2,2,0)$ supermultiplets (for target dimensions $D=1,2$, respectively). One sign of the obstruction for $D=3$ is the appearance in the Hamiltonian of the non-diagonal spin-orbit term.\par 
The third approach to superconformal quantum mechanics is based on symmetries of the time-dependent 
Schr\"odinger equation, regarded as a partial differential equation (it will be briefly discussed in the Conclusions).
\par
We mention that several different, both classical and quantum, supersymmetric models possessing (a real form of)  $sl(2|1)$ as dynamical
symmetry have been investigated in the literature, see \cite{{dv},{su21is}, {cuhoto},{su21ist},{su21iva}, {su21fils}, {su21koz}}.  These models do not correspond to the three-dimensional Hamiltonians here presented. The first one of these papers, \cite{dv}, presents the $sl(2|1)$ dynamical symmetry of a Dirac magnetic monopole with a $\frac{1}{r^2}$ potential.\par
Hamiltonians with a spin-orbit coupling, as the one here discussed, are not a novelty. It has to be mentioned, in particular, the \cite{bal} paper. In that work a particular $2\times 2$ matrix Hamiltonian with spin-orbit coupling
has been solved by showing that the system possesses an $osp(1|2)$ dynamical symmetry. The restriction to the upper left block of the
$4\times 4$ matrix Hamiltonian (\ref{hamosc}) derived below produces, at the specific $\beta=\frac{1}{2}$ value, the
Hamiltonian in formula (2.6) of \cite{bal} (a further constant $\lambda$, entering the (2.6) Hamiltonian, is set to zero).  Since the $\beta$ coupling constant entering \cite{bal} is kept fixed, from that work no features can be derived concerning the zigzag pattern of vacuum energy or the interpolating regimes obtained at varying $\beta$.
\par
The scheme of the paper is the following. In Section {\bf 2} we introduce the three-dimensional superconformal
quantum mechanics. In Section {\bf 3} we apply the de Alfaro-Fubini-Furlan ``trick" to construct the associated
deformed oscillator with $sl(2|1)$ spectrum-generating superalgebra. The selection of its admissible Hilbert spaces is discussed in Section {\bf 4}.  The derivation of its spectrum is given in Section {\bf 5}. The computation of the energy degeneracy is presented in Section {\bf 6}. The recovering of previous models from dimensional reduction is explained in Section {\bf 7}. In the Conclusions we mention open problems and lines of future research. The paper is complemented by three Appendices. Our notations of conventions are introduced in Appendix {\bf A}. In Appendix {\bf B} the orthonormal eigenstates are computed. We illustrate in Appendix {\bf C}
the open problem of the reducibility of the Hilbert space with respect to the $sl(2|1)$ lowest weight representations versus its possible irreducibility with respect to a representation of a larger algebraic structure. 

\section{The three-dimensional superconformal quantum mechanics}

We realize, following the approaches in \cite{akt} and \cite{ackt}, a superconformal dynamical symmetry in terms
of first-order matrix differential operators. Several requirements have to be satisfied. The operators have to be Hermitian. The fermionic ones need to be block-antidiagonal in order to be accommodated into the odd sector
of the superalgebra. A supersymmetric quantum mechanics has to be constructed at first. For our purposes the
supersymmetry generators have to be the square roots of a Hamiltonian which corresponds to a deformation of the 
three-dimensional Laplacian of the free theory. It is easily seen that a three-dimensional Laplacian is nicely expressed
in terms of quaternions (which require at least $4\times 4$ {\em real} matrices).  On the other hand, as recalled in \cite{ackt}, the closure
with Hermitian operators of
a superconformal dynamical symmetry (which contains in particular the conformal partner of the Hamiltonian), requires the introduction of complex matrices. All in all, $4\times 4$ complex matrices is the minimal set-up to achieve the goal (it follows from the properties of Clifford algebras discussed, e.g., in \cite{oku} and \cite{crt}). It allows to produce an ${\cal N}=2$ extended supersymmetric quantum mechanics, due to the presence of two block-antidiagonal matrices $\gamma_a$ ($a=1,2$) which commute with the three imaginary quaternions $h_i$ ($i=1,2,3)$.
An explicit realization of the matrices $\gamma_a$, $h_i$ and of the Fermion Parity Operator $N_F=\gamma_3$ is given in (\ref{4x4matrices}).
The two supersymmetry operators $Q_a$ are further assumed to have scaling dimension $[Q_a]=\frac{1}{2}$ if
$[x_i]=-\frac{1}{2}$ is the assigned scaling dimension of the three space coordinates $x_i$ (our notations and conventions are given in Appendix {\bf A}). \par
A natural Ansatz to produce two supersymmetry operators $Q_a$ with the required properties consists in setting
\bea\label{qaop}
Q_a &=&\frac{1}{\sqrt 2}
\gamma_{a}\left({\slashed{\partial}}-\frac{\beta}{r^2}N_F{\slashed{r}}\right).
\eea
In the above formula $\beta$ is a real parameter, $r=\sqrt{x_1^2+x_2^2+x_3^2}$ is the radial coordinate, while $
{\slashed{\partial}} = \partial_ih_i$ and ${\slashed{r}}=x_ih_i$ are introduced in (\ref{rhLh}).\par
The superalgebra of the ${\cal N}=2$ supersymmetric quantum mechanics is
\bea
\{ Q_a, Q_b\} = 2\delta_{ab} H,\quad &&\quad [H, Q_a]=0.
\eea 
The $4\times 4$ matrix supersymmetric Hamiltonian $H$ is given by
\bea\label{hop}
H
 & =&\left(\begin{array}{cc}
(-\frac{1}{2}\nabla^{2}+\frac{2\beta}{r^{2}}{\overrightarrow{{\mathbf{S}}}}\cdot{\overrightarrow{\mathbf{L}}}+\frac{\beta\left(\beta+1\right)}{2r^{2}}){\mathbb I}_2 & 0\\
0 & (-\frac{1}{2}\nabla^{2}-\frac{2\beta}{r^{2}}{\overrightarrow{{\mathbf{S}}}}\cdot{\overrightarrow{\mathbf{L}}}+\frac{\beta\left(\beta-1\right)}{2r^{2}}){\mathbb I}_2
\end{array}\right),
\eea
where $\nabla^{2}=\partial_{x_1}^2+\partial_{x_2}^2+\partial_{x_3}^2$ is the three-dimensional Laplacian
and ${\overrightarrow{\mathbf S}}$ is the spin-$\frac{1}{2}$ introduced in (\ref{spin}).
At $\beta=0$ the Hamiltonian gets reduced to $H=-\frac{1}{2}\nabla\cdot {\mathbb I}_4$. At non-vanishing $\beta$ two
extra terms appear: a $\frac{1}{r^2}$ diagonal potential term proportional to $\beta(\beta\pm 1)$ in the upper/lower diagonal blocks
and a non-diagonal spin-orbit interaction term proportional to ${\overrightarrow{{\mathbf{S}}}}\cdot{\overrightarrow{\mathbf{L}}}$. The latter one is a first-order differential operator.\par
The Hamiltonian $H$ has scaling dimension $[H]=1$. Based on the de Alfaro-Fubini-Furlan construction \cite{dff},
we can introduce its conformal partner as the rotationally invariant operator $K$ of scaling dimension $[K]=-1$.
We can therefore set
\bea\label{kop}
K&=& \frac{1}{2}r^2{\mathbb I}_4
\eea
and verify whether the repeated (anti)commutators of the operators $Q_a$ and $K$ close the ${\cal N}=2$ one-dimensional superconformal algebra $sl(2|1)$ (see \cite{{tian},{ackt}} for a discussion of one-dimensional, ${\cal N}$-extended  superconformal algebras). This is indeed the case. Four extra operators (${\overline Q}_a, D, R$) have to be added. $D$ is the (bosonic) dilatation operator which, together with $H,K$, close the $sl(2)$ subalgebra. The two fermionic operators ${\overline Q}_a$, of scaling dimension $[{\overline Q}_a]=-\frac{1}{2}$, are introduced from the commutators $[Q_a,K]$.
Finally, $R$ is the $u(1)$ $R$-symmetry bosonic operator of $sl(2|1)$. It is introduced from the anticommutators $\{Q_a, {\overline Q}_b\}$ with $a\neq b$. The (anti)commutators among the eight operators $H,D,K,R, Q_a, {\overline Q}_a$ close the $sl(2|1)$ superalgebra. We present them for completeness. The non-vanishing ones are
\bea\label{sl21comanticom}
&\begin{array}{clclcl}[D,H]&=-2iH,&[D,K]&=2iK,&[H,K]&= iD,\\
\relax [D,Q_a]&=-iQ_a,&[D,{\overline Q}_a]&=i{\overline Q}_a,&&\\
\relax [H,{\overline Q}_a]&= iQ_a&[K,{ Q}_a]&=-i{\overline Q}_a,&&\\
\{Q_a,Q_b\}&=2\delta_{ab}H,&\{{\overline Q}_a,{\overline Q}_b\}&=2\delta_{ab}K,&\{Q_a,{\overline Q}_b\}&=\delta_{ab}D+\epsilon_{ab}R,\\
\relax [R,Q_a]&=-3i\epsilon_{ab}Q_b,&[R,{\overline Q}_a]&=-3i\epsilon_{ab}{\overline{Q}}_b,&&
\end{array}&
\eea
with the antisymmetric tensor $\epsilon_{ab}$ normalized so that $\epsilon_{12}=1$.\par
Besides $Q_a$, $H$, $K$, respectively given in (\ref{qaop},\ref{hop},\ref{kop}), the remaining operators are
\bea\label{remop}
{\overline Q}_a&=&-\frac{i}{\sqrt{2}}\gamma_a{\slashed{r}},\nonumber\\
D&=&i (x_j\partial_j+\frac{3}{2})\cdot {\mathbb I}_4= i(r\partial_r+\frac{3}{2})\cdot{\mathbb I}_4,\nonumber\\
R&=&-(\frac{3}{2}N_F+\beta\cdot {\mathbb I}_4).
\eea
The Hamiltonian $H$ is, by construction, Hermitian. Since the spin is $\frac{1}{2}$, the total angular 
momentum ${\overrightarrow {\bf J}}={\overrightarrow{\bf L}}+{\overrightarrow{\bf S}}$ of the quantum-mechanical system is half-integer. The Hamiltonian is non-diagonal; on the other hand, due to the relation
\bea\label{spinorbitop}
{\overrightarrow {\bf L}}\cdot{\overrightarrow {\bf S}}&=& \frac{1}{2}({\overrightarrow {\bf J}}^2-{\overrightarrow{\bf L}}^2-{\overrightarrow{\bf S}}^2)
= \frac{1}{2}(j(j+1)-l(l+1)-\frac{3}{4}),
\eea
it gets diagonalized in each sector of given total $j$ and orbital $l$ angular momentum. In each such sector 
it corresponds to a constant kinetic term plus a diagonal potential term proportional to $\frac{1}{r^2}$.

\section{The three-dimensional deformed oscillator}

By setting, following \cite{dff},
\bea
H_{osc}&=& H+K,
\eea
with $H$, $K$ respectively given in (\ref{hop}) and (\ref{kop}), we produce the $4\times 4$ matrix deformed oscillator Hamiltonian $H_{osc}$ whose spectrum is discrete and bounded from below. By construction, the $sl(2|1)$
dynamical symmetry of the $H$ Hamiltonian acts as a spectrum-generating superalgebra for the $H_{osc}$ Hamiltonian.\par
The explicit expression of $H_{osc}$ is
\bea\label{hamosc}
H_{osc} &=& -\frac{1}{2}\nabla^2\cdot{\mathbb I}_4+\frac{1}{2r^2}(\beta^2\cdot{\mathbb I}_4+ \beta N_F(1+4\cdot{\mathbb I}_2\otimes{\vec{\bf S}}\cdot{\vec {\bf L}}))+\frac{1}{2}r^2\cdot {\mathbb I}_4.
\eea
The spin of the $H_{osc}$ quantum mechanical system is $\frac{1}{2}$. Therefore, its total angular momentum $j$ is half-integer ($j\in\frac{1}{2}+{\mathbb N}_0$) and the relation with the orbital angular momentum $l\in{\mathbb N}_0$ is
given by
\bea
j= l+\delta\frac{1}{2},&\quad&\textrm{for}\quad \delta=\pm 1.
\eea
In the given $j,l$ sector, the operator ${\vec{\bf L}}\cdot {\vec{\bf S}}$ from (\ref{spinorbitop}) is constant. We get
\bea
{\vec{\bf L}}\cdot {\vec{\bf S}} =\frac{1}{2}\alpha, &\quad& \textrm{with} \quad \alpha=\delta (j+\frac{1}{2})-1.
\eea
Each given bosonic (or fermionic) energy eigenvalue in the $j,l$ sector is $(2j+1)$-times degenerated, the degenerate
eigenstates being labeled by the $J_z\equiv J_3$ quantum numbers $-j, j-1,\ldots, j$.\par
The energy eigenstates of the system are described with the help of the two-component $\mathcal{Y}_{j,l,m}\left(\theta,\phi\right)$ spin spherical harmonics, see \cite{bieden},
given by
\bea\label{spinspherical}
\mathcal{Y}_{j,j-\frac{1}{2}\delta,m}\left(\theta,\phi\right) & =& \frac{1}{\sqrt{2j-\delta+1}}\left(\begin{array}{c}
\delta\sqrt{j+\frac{1}{2}(1-\delta)+\delta m}Y_{j-\frac{1}{2}\delta}^{m-\frac{1}{2}}\left(\theta,\phi\right)\\
\sqrt{j+\frac{1}{2}(1-\delta)-\delta m}Y_{j-\frac{1}{2}\delta}^{m+\frac{1}{2}}\left(\theta,\phi\right)
\end{array}\right),
\eea 
where $Y_{l}^{n}(\theta,\phi)$ (for $n=-l,-l+1,\ldots, l$) are the ordinary spherical harmonics.\par
The spin spherical harmonics $\mathcal{Y}_{j,j-\frac{1}{2}\delta,m}\left(\theta,\phi\right)$ are the eigenstates for the compatible
observable operators ${\vec{\bf J}}\cdot{\vec{\bf J}}$, ${\vec{\bf L}}\cdot{\vec{\bf L}}$, $J_z$, with eigenvalues
$j(j+1)$, $(j-\frac{1}{2}\delta)(j-\frac{1}{2}\delta+1)$, $m$, respectively.
\par
The spectrum of the model is derived from the creation (annihilation) operators $a_b^\dagger$ ($a_b$), with $b=1,2$, which are introduced through
the positions
\bea
a_b =Q_b+i{\overline Q}_b, &\quad& a_b^\dagger=Q_b-i{\overline Q}_b.
\eea
Indeed, we obtain
\bea
H_{osc} &=&\frac{1}{2}\{a_1,a_1^\dagger\}=\frac{1}{2}\{a_2,a_2^\dagger\},
\eea
together with
\bea
[H_{osc},a_b]=-a_b, &\quad&[H_{osc},{a_b}^\dagger]=a_b^\dagger.
\eea
For completeness we also present the commutators
\bea\label{cmmut}
[a_1,a_1^\dagger]=[a_2,a_2^\dagger]&=& 3\cdot {\mathbb I}_4 +4\cdot {\mathbb I}_2\otimes {\vec {\bf S}}\cdot {\vec {\bf L}}-2\beta N_F.
\eea
They should be compared with the analogous commutators for one-dimensional deformed oscillators (see \cite{ackt}),
producing deformed Heisenberg algebras with diagonal operators in the right hand side; the right hand side of (\ref{cmmut}) is diagonalized in the $j,l$ sector.\par
Let us set, for convenience,  $a_b^-\equiv a_b$ and $a_b^+\equiv a_b^\dagger$. The explicit expression of the creation/annihilation operators is
\bea\label{creation}
a_b^\pm &=&\frac{\slashed{r}}{r\sqrt 2 }\gamma_b({\mathbb I}_4\cdot(\partial_r\mp r)-\frac{2}{r}{\mathbb I}_2\otimes 
{\vec{\bf S}}\cdot{\vec{\bf L}}-\frac{\beta}{r}N_F).
\eea
They can be factorized as
\bea\label{factor}
a_b^\pm = \frac{\slashed{r}}{r\sqrt 2 }\gamma_ba^\pm, &\quad& \textrm{with} \quad a^\pm=({\mathbb I}_4\cdot(\partial_r\mp r)-\frac{2}{r}{\mathbb I}_2\otimes 
{\vec{\bf S}}\cdot{\vec{\bf L}}-\frac{\beta}{r}N_F).
\eea
The creation/annihilation operators $a_b^\pm$ anticommute with the Fermion Parity Operator $N_F$,
\bea
\{a_b^\pm,N_F\}&=&0,
\eea
producing towers of alternating bosonic/fermionic energy eigenstates.\par
A lowest weight state $\Psi_{lws}$ is defined to satisfy 
\bea\label{lws}
a_b^- \Psi_{lws}&=& 0.
\eea
Due to the (\ref{factor}) factorization, in both $b=1,2$ cases, this is tantamount to satisfy
$a^-\Psi_{lws}=0$.\par
A lowest weight representation is spanned by the action of the $a_b^+$ creation operators on $\Psi_{lws}$.\par
If $\Psi_{lws}$ is either bosonic or fermionic, the vectors $a_1^+v$ and $a_2^+v$, with $v$ belonging to the lowest weight representation, differ by a phase. Therefore, the action of $a_1^+$, $a_2^+$ produces the same ray vector characterizing a physical state of the Hilbert space. This phenomenon was already observed \cite{ackt} in the one-dimensional
context. Without loss of generality we can just pick, let's say, $a_1^\pm$ to create/annihilate the ray vectors. \par
We search for solutions $\Psi_{j,\delta,m}^\epsilon(r,\theta,\phi)$ of the lowest weight condition (\ref{lws}) of the form 
\bea\label{lwv}
\Psi_{j,\delta,m}^\epsilon(r,\theta,\phi)&=& f_{j,\delta}^\epsilon (r)\cdot e_{\epsilon}\otimes \mathcal{Y}_{j,j-\frac{1}{2}\delta,m}\left(\theta,\phi\right),\quad\textrm{with}\quad \epsilon =\pm 1.
\eea
The sign of $\epsilon$ (no summation over this repeated index) refers to the bosonic (fermionic) states with respective eigenvalues $\epsilon =+1$ ($\epsilon=-1$) of the Fermion Parity Operator $N_F$; we have $e_{+1}=\tiny{\left(\begin{array}{c}1\\ 0\end{array}\right)}$ and $ e_{-1}={\tiny{\left(\begin{array}{c}0\\ 1\end{array}\right)}}$. \par
Solutions of (\ref{lws}) are obtained for the radial-coordinate functions $f_{j,\delta}^\epsilon (r)$ given by
\bea \label{radialf}
f_{j,\delta}^\epsilon (r)&=&r^{\gamma_{(j,\delta,\epsilon)}}e^{-\frac{1}{2}r^2},
\eea
where
\bea\label{gamma}
{\gamma_{(j,\delta,\epsilon)}}(\beta)&=&\alpha+\beta\epsilon=\delta(j+\frac{1}{2})+\beta\epsilon-1.
\eea
The corresponding lowest weight state energy eigenvalue $E_{j,\delta,\epsilon}(\beta)$ from
\bea
H_{osc}(\beta)\Psi_{j,\delta,m}^\epsilon(r,\theta,\phi)&=&E_{j,\delta,\epsilon}(\beta)\Psi_{j,\delta,m}^\epsilon(r,\theta,\phi)
\eea
is
\bea \label{energy}
E_{j,\delta,\epsilon}(\beta)&=&\delta(j+\frac{1}{2})+\beta\epsilon+\frac{1}{2}.
\eea
Since $E_{j,\delta,\epsilon}(\beta)$ does not depend on the quantum number $m$, this energy eigenvalue is 
$(2j+1)$ times degenerate.\par
We discuss in the next Section under which condition the vectors of a given lowest weight representation
belong to a normed Hilbert space.  For the time being we point out that the application of the creation operator $a_1^\dagger$ on an energy eigenstate with energy eigenvalue $E$ produces a ray vector with energy eigenvalue $E+1$.  As we will see, an admissible Hilbert space is defined by a direct sum of (an infinite number of) $sl(2|1)$ lowest weight representations. The degeneracy of each energy level is finite and can be computed with a recursive formula. Let $n(E)$ be the total number of distinct, admissible, lowest weight vectors in the Hilbert space and let $d(E)$ be the number of degenerate eigenstates at energy level $E$. At energy level $E+1$ we obtain
\bea\label{iterative}
d(E+1)&=& d(E) +n(E+1).
\eea
The $d(E)$  term in the right hand side gives the number of descendant states obtained by applying $a_1^\dagger$ to the degenerate states at energy $E$, while the $n(E+1)$ term corresponds to the number of
new primary states at $E+1$. From (\ref{iterative}), the computation of the degeneracy of the spectrum is reduced to a combinatorial problem based on the determination of the lowest weight vectors.

\section{Alternative Hilbert spaces for the $3D$ deformed oscillator}

Without loss of generality we can restrict the real parameter $\beta$ to belong to the half-line $\beta\geq 0$
since the mapping $\beta\leftrightarrow -\beta$ is recovered by a similarity transformation (induced by an operator $S$) which exchanges bosons
into fermions. We have, indeed,
\bea
S H_{osc}(\beta)S^{-1} &=& H_{osc}(-\beta) \quad\quad\textrm{with}\quad S=\sigma_1\otimes{\mathbb I}_2.
\eea
At $\beta=0$ we recover four copies of the undeformed, isotropic, three-dimensional oscillator. Therefore, $\beta >0$ parametrizes the deformed oscillator.\par 
To the following $j,\delta,\epsilon, m$ quantum numbers, 
\bea\label{quantumnumbers}
&j\in \frac{1}{2}+{\mathbb N}_0, \quad\quad \delta =\pm 1,\quad\quad\epsilon=\pm 1,\quad\quad m=-j,-j+1,\ldots , j,
\eea
is associated an $sl(2|1)$ lowest weight vector and its induced representation.\par
At a given $\beta$, the Hilbert space has to be selected by requiring, in particular, the normalization of the wave functions entering a lowest weight representation. This puts restriction on the admissible lowest weight representations, the Hilbert space being constructed as the direct sum of the admissible lowest weight representations.  \par
The analysis of \cite{{mt},{ftf}} (and also of \cite{ackt}) can be extended to the present case. Two choices to select the Hilbert space naturally appear:
\begin{itemize}
\item {\em case ~ i}: ~ the wave functions can be singular at the origin, but they need to be normalized,
\item {\em case  ~ii}: ~ the wave functions are assumed to be regular at the origin.
\end{itemize}
We will see in a moment that {\em case i} corresponds in restricting the admissible lowest weight representations
to those satisfying the necessary and sufficient condition
\bea\label{normal1}
2\gamma_{(j,\delta,\epsilon)}(\beta)+3&>&0,
\eea
where $\gamma_{(j,\delta,\epsilon)}(\beta)$ was introduced in (\ref{radialf}) and given in (\ref{gamma}). The normalizability condition (\ref{normal1}) is equivalent to the requirement
\bea\label{normal2}
E_{j,\delta,\epsilon}(\beta)&>&0
\eea
for the lowest weight energy $
E_{j,\delta,\epsilon}(\beta)$ given in (\ref{energy}).\par
The {\em case ii} (regularity at the origin) corresponds in restricting the admissible lowest weight representations
to those satisfying the condition
\bea\label{regull1}
\gamma_{(j,\delta,\epsilon)}(\beta)&\geq&0 \quad\quad \textrm{for}\quad \beta\geq 0.
\eea
The single-valuedness of the wave functions at the origin implies that the equality 
$\gamma_{(j,\delta,\epsilon)}(\beta)=0$ can only be realized with vanishing ($l=0$) orbital angular momentum
(therefore, for $\delta=1$ and $j=\frac{1}{2}$). At $\beta=0$ one recovers the vacuum state of the undeformed oscillator. For the deformed $\beta>0$ oscillator the strict inequality
\bea\label{regul2}
\gamma_{(j,\delta,\epsilon)}(\beta)&>&0\quad \quad \textrm{for} \quad \beta>0
\eea
is required as necessary and sufficient condition.\par
For illustrative purposes it is useful to present a table (up to $j=\frac{5}{2}$) of the $\beta$ range of admissible lowest weight representations under {\em norm} ({\em case ~i}) and {\em reg} ({\em case ~ii}) conditions. The $j,\delta,\epsilon$ quantum numbers are respectively given in columns $1,2,3$, while $\gamma_{(j,\delta,\epsilon)}(\beta)$ and $ E_{j,\delta,\epsilon}(\beta)$ are presented in columns $4$ and $5$. We have

\bea
&\begin{array}{|c|c|c||c|c||c|c|}\hline
j&\delta&\epsilon&\gamma&E&norm&reg\\ \hline\hline
\frac{1}{2}&+&+&\beta&\frac{3}{2}+\beta&\beta\geq 0&\beta\geq 0\\ \hline
\frac{1}{2}&+&-&-\beta&\frac{3}{2}-\beta&0\leq \beta<\frac{3}{2}&\beta=0\\ \hline
\frac{1}{2}&-&+&\beta-2&-\frac{1}{2}+\beta&\beta>\frac{1}{2}&\beta>2\\ \hline
\frac{1}{2}&-&-&-\beta-2&-\frac{1}{2}-\beta& \times&\times\\ \hline\hline
\frac{3}{2}&+&+&\beta+1&\frac{5}{2}+\beta&\beta\geq 0&\beta\geq 0\\ \hline
\frac{3}{2}&+&-&-\beta+1&\frac{5}{2}-\beta&0\leq \beta<\frac{5}{2}&0\leq\beta<1\\ \hline
\frac{3}{2}&-&+&\beta-3&-\frac{3}{2}+\beta&\beta>\frac{3}{2}&\beta>3\\ \hline
\frac{3}{2}&-&-&-\beta-3&-\frac{3}{2}-\beta& \times&\times\\ \hline\hline
\frac{5}{2}&+&+&\beta+2&\frac{7}{2}+\beta&\beta\geq 0&\beta\geq 0\\ \hline
\frac{5}{2}&+&-&-\beta+2&\frac{7}{2}-\beta&0\leq \beta<\frac{7}{2}&0\leq \beta<2\\ \hline
\frac{5}{2}&-&+&\beta-4&-\frac{5}{2}+\beta&\beta>\frac{5}{2}&\beta>4\\ \hline
\frac{5}{2}&-&-&-\beta-4&-\frac{5}{2}-\beta& \times&\times\\ \hline
\end{array}
&
\eea
For the undeformed $\beta=0$ oscillator, the restrictions from either {\em case  i} or {\em case ii} select the same Hilbert space. It is the direct sum of all lowest weight representations with $j,\epsilon, m$ satisfying (\ref{quantumnumbers}),
while $\delta$ is restricted to be
\bea
\delta&=&+1.
\eea
For the $\beta>0$ deformed oscillators,  the Hilbert spaces ${\cal H}_{norm}$ and ${\cal H}_{reg}$ are direct sums of the lowest weight representations with $j\in\frac{1}{2}+{\mathbb N}_0$ satisfying (depending on $\delta$, $\epsilon$)
\bea\label{jadmissible}
&\begin{array}{|c|c|c|c|}\hline&&{\cal H}_{norm}:&{\cal H}_{reg}:\\ \hline
\delta=+1&\epsilon=+1&\textrm{any}~j&\textrm{any}~j\\ \hline
\delta=+1&\epsilon=-1&j>\beta-1&j>\beta+\frac{1}{2}\\ \hline
\delta=-1&\epsilon=+1&j<\beta&j<\beta-\frac{3}{2}\\ \hline
\delta=-1&\epsilon=-1&\textrm{no}~j&\textrm{no}~j\\ \hline
\end{array}&
\eea
The vacuum energy of the system is obtained by comparing the smallest lowest weight energies (obtained from (\ref{energy})) in each one of the three contributing sectors ($\delta=1$ with $\epsilon=1$,  $\delta=1$ with $\epsilon=-1$, $\delta=-1$ with $\epsilon=1$) for the admissible values of $j$. One should take into account that for
$\delta=+1$, at a given $\epsilon=\pm 1$, the smallest value of energy is encountered at the smallest admissible value of $j$. Conversely,
for $\delta=-1$, $\epsilon=+1$, the smallest lowest weight energy is encountered at the largest admissible value of $j$.\par
The normalizability condition (\ref{normal1}) comes from the requirement to have a finite norm for the lowest weight vector (\ref{lwv}), such that
\bea
\int d^3x Tr\left({\Psi_{j,\delta,m}^\epsilon(r,\theta,\phi)}^\dagger
\Psi_{j,\delta,m}^\epsilon(r,\theta,\phi)\right)&<&\infty.
\eea
Passing to spherical coordinates, the only potentially troublesome integral is
\bea\label{gammafromhere}
&\int_0^{+\infty}dr r^2 r^{2\gamma_{(j,\delta,\epsilon)}}e^{-r^2},&
\eea
which is finite at the origin, provided that ${2\gamma_{(j,\delta,\epsilon)}}+2>-1$,  namely the (\ref{normal1}) condition. A simple inspection shows that, acting with the $a_1^+$ creation operators, no further condition is 
necessary to ensure the normalizability of the excited states of the lowest weight representation.\par
The normalizability condition discussed in the \cite{{mt},{ftf},{ackt}} papers (which deal with one-dimensional systems) implies that the
normalized wave functions are defined on the real line, with the possible ``smooth" singularity at the origin. The regular wave functions (for $\beta >0$) are, on the other hand, defined on the half line and satisfy the Dirichlet boundary condition at the origin. We stress that the mathematical framework in the present work coincides with
the one introduced in \cite{{mt},{ftf},{ackt}}. For the singular Hilbert space ${\cal H}_{norm}$ this means that one
can extend the wavefunctions (defined by the spherical coordinates $r,\theta,\phi$) in two regions, namely
$I_+$ (parametrized by $r\geq 0$) and $I_-$ (parametrized by $r\leq 0$), so that their intersection
$I_+\cap I_- = \{O\}$ is the origin. Mathematically, this construction works fine. On a physical ground one has to reconcile the interpretation of the wavefunctions defined on $I=I_+\cup I_-$ with the wavefunctions defined on the ordinary
three-dimensional space. This can be achieved by taking into account that the ${\bf Z}_2$ transformation
\bea
Z&:& r\mapsto  -r\quad ( \textrm{with}\quad \theta,\phi\quad\textrm{unchanged}\quad\textrm{and}\quad Z^2=1),
\eea
is a symmetry of the $H_{osc}$ Hamiltonian.  The induced operator ${\widehat Z}$, acting on the  wavefunctions
defined on $I_+\cup I_-$, satisfies the
\bea\label{z2identity}
{\widehat Z}^2&=& {\mathbb I}
\eea
condition, is Hermitian and possesses $\pm 1$ eigenvalues, provided that 
\bea \label{lwvcondition}
(-1)^{\gamma_{(j,\delta,\epsilon)}} &=&\pm 1,
\eea
implying that $\gamma_{(j,\delta,\epsilon)}(\beta)$ (and consequently $\beta$) is an integer.
If (\ref{lwvcondition}) is satisfied for the lowest weight vector, ${\widehat Z}^\dagger={\widehat Z}$ and
(\ref{z2identity}) are satisfied for all states belonging to the associated lowest weight representation. For such a well-defined ${\widehat Z}$ operator, one can impose the $\{I_+\cup I_-\}\slash {\bf Z}_2$ coset construction, 
by identifying
\bea
(r,\theta,\phi) &\equiv & (-r,\theta,\phi).
\eea
Under this coset construction, one can set
\bea\label{coset}
\{I_+\cup I_-\}\slash {\bf Z}_2 &\approx& {\mathbb R}^3.
\eea
The ${\bf Z}_2$ coset construction entails, from equations (\ref{lwvcondition}) and (\ref{gamma}), the quantization
of $\beta$ which, under this condition, is restricted to be integer
\bea\label{quantization}
\beta &\in & {\mathbb Z}.
\eea
We stress the fact that the Hilbert space can be consistently defined for any real value of $\beta$. It is the ${\bf Z}_2$ coset construction with the associated interpretation which requires $\beta$ to be an integer.

\section{The spectrum of the $\beta$-deformed oscillator}

In this Section we present, for $H_{osc}$, the computation of the spectrum for the alternative choices of ${\cal H}_{norm}$ and ${\cal H}_{reg}$ Hilbert spaces. Without loss of generality $\beta$ is taken to be $\beta\geq 0$. We start with the ${\cal H}_{norm}$ case (normalized wave functions).
\subsection{The spectrum for the ${\cal H}_{norm}$ Hilbert space of normalized wave functions}
For $\beta\geq \frac{1}{2}$ it is convenient to introduce, via the floor function, the parameter $\mu$, defined as
\bea
&\mu = \{\beta-\frac{1}{2}\} =(\beta-\frac{1}{2})-\lfloor \beta-\frac{1}{2}\rfloor, \quad\quad  p=\lfloor \beta-\frac{1}{2}\rfloor,&\nonumber\\
&\textrm{so that} \quad \mu\in [0,1[,\quad p\in {\mathbb N}_0 \quad \textrm{and}\quad~~~~~ \beta=\frac{1}{2}+\mu+p.&
\eea
The results for the spectrum split into six different cases which have to be separately analyzed:
\begin{itemize}
\item {\bf case ~ I}: ~ $\beta=0$ (the undeformed oscillator),
\item {\bf case  ~II}: ~ $\beta= 1+p$, with $p\in {\mathbb N}_0$ ($p=0,1,2,\ldots $),
\item {\bf case III}: ~ $\beta=\frac{1}{2}+p$, with $p\in{\mathbb N}_0$,
\item {\bf case IV}: ~ $0<\beta<\frac{1}{2}$,
\item {\bf case ~V}: ~ $0<\mu<\frac{1}{2}$, therefore  $\beta=\frac{1}{2}+\mu+p$, with $p\in{\mathbb N}_0$,
\item {\bf case VI}: ~ $\frac{1}{2}<\mu<1$, therefore $\beta=\frac{1}{2}+\mu+p$, with $p\in{\mathbb N}_0$.\bea
&&
\eea
\end{itemize}
The energy eigenvalues corresponding to the above cases are
\begin{itemize}
\item {\bf case ~ I}: ~ $E_n=\frac{3}{2}+n$, where $n\in {\mathbb N}_0$ is a non-negative integer.\\
The vacuum energy is $E_{vac}=\frac{3}{2}$; the ground state is four times degenerated, with two bosonic and two fermionic eigenstates (hence ``$2_B+2_F$").\\
The vacuum lowest weight vectors are specified by the quantum numbers $j=\frac{1}{2}$, $\delta=+1$, $\epsilon=\pm 1$ and (here and in the following) all compatible values $m=-j,\ldots, j$.
\item {\bf case  ~II}:  ~ $E_n=\frac{1}{2}+n$, with $n\in {\mathbb N}_0$.\\
The vacuum energy is $E_{vac}=\frac{1}{2}$; the degeneration of the ground state is $2(p+1)$, with $p+1$ bosonic and $p+1$ fermionic eigenstates, and is therefore denoted as ``$(p+1)_B+(p+1)_F$".\\
The vacuum lowest weight vectors are specified by $j=\frac{1}{2}+p$, with either $\delta=+1$, $\epsilon=-1$ or
$\delta=-1$, $\epsilon=+1$.
\item {\bf case III}:  ~ $E_n=1+n$, with $n\in {\mathbb N}_0$.\\
The vacuum energy is $E_{vac}=\frac{1}{2}$; 
 the degeneration of the ground state is $4p+2$, with $2p$ bosonic and $2(p+1)$ fermionic eigenstates, and is therefore denoted as ``$(2p)_B+(2p+2)_F$".
\\
For $p=0$ the two vacuum lowest vectors are specified by $j=\frac{1}{2}$, $\delta=+1$, $\epsilon = -1$. \\For $p>0$
the vacuum lowest vectors are specified either by $j=\frac{1}{2}+p$, $\delta= +1$, $\epsilon =-1$ or by
$j=p-\frac{1}{2}$, $\delta=-1$, $\epsilon=+1$.
\item {\bf case IV}: two series of energy eigenvalues $E_n^\pm= \frac{3}{2}\pm \beta+n$, with $n\in {\mathbb N}_0$, are encountered.\\
The vacuum energy is $E_{vac}=\frac{3}{2}-\beta$;
the ground state is fermionic and doubly degenerated (``$2_F$"). \\
The two vacuum lowest weight vectors are specified by $j=\frac{1}{2}$, $\delta=+ 1$, $\epsilon=- 1$.
\item {\bf case ~V}:  two series of energy eigenvalues $E_n^-= \mu+n$, $E_n^+=1- \mu+n$, with $n\in {\mathbb N}_0$, are encountered.\\
The vacuum energy is $E_{vac}=\mu$; the ground state is bosonic and $(2p+2)$-times degenerated (hence ``$(2p+2)_B$"). \\
The vacuum lowest weight vectors are specified by $j=\frac{1}{2}+p$, $\delta=- 1$, $\epsilon=+ 1$.
\item {\bf case VI}: two series of energy eigenvalues $E_n^-= 1-\mu+n$, $E_n^+=\mu+n$, with $n\in {\mathbb N}_0$, are encountered.\\
The vacuum energy is $E_{vac}=1-\mu$; the ground state is fermionic and $(2p+2)$-times degenerated (hence ``$(2p+2)_F$"). \\
The vacuum lowest weight vectors are specified by $j=\frac{1}{2} +p$, $\delta=+1$, $\epsilon=- 1$.\bea
&&\label{normcases}
\eea
\end{itemize}
There is an important remark. The energy spectrum of the {\bf V} and {\bf VI} cases coincides under a
\bea
\mu&\leftrightarrow 1-\mu, \quad\quad \textrm{with} \quad \mu\neq0,\frac{1}{2},
\eea
duality transformation. Under this duality transformation the parity (bosonic/fermionic) of the ground state is exchanged.
On the other hand, the degeneracies of the energy eigenvalues above the ground state are not respected by the duality transformation. It is sufficient to discuss a specific example to show it. Let's take $\mu=\frac{1}{4}$ with $p=0$ and consider the dually related $\beta=\frac{3}{4}$ and $\beta=\frac{5}{4}$ cases.\par
The lowest weight vectors appearing in the first five energy levels (from the vacuum energy $E_{vac}=\frac{1}{4}$ to $E=\frac{9}{4}$) are presented in the following table
\bea
&\begin{array}{|c||c||c|}\hline
E&\beta=\frac{3}{4}&\beta=\frac{5}{4}\\ \hline\hline
\frac{9}{4}&\frac{1}{2}+B&\frac{5}{2}+F\\ \hline
\frac{7}{4}&\frac{3}{2}+F&\times\\ \hline
\frac{5}{4}&\times&\frac{3}{2}+F\\ \hline
\frac{3}{4} &\frac{1}{2}+F&\frac{1}{2}-B\\ \hline
\frac{1}{4}&\frac{1}{2}-B&\frac{1}{2}+F\\ \hline
\end{array}
&
\eea
The lowest weight vectors are identified by their quantum numbers $j$, the $\pm$ sign referring to $\delta$, while $B$ (standing for bosons) and $F$ (standing for fermions) correspond to $\epsilon=+1$, $\epsilon=-1$, respectively.\par

The degeneracy $d_\beta(E)$ of a given energy level is computed with the recursive formula (\ref{iterative}) involving
primary (lowest weight vectors) and descendant states. The results are presented in the table
\bea
&\begin{array}{|c||c||c|}\hline
E&d_{\beta=\frac{3}{4}}(E)&d_{\beta=\frac{5}{4}}(E)\\ \hline\hline
\frac{9}{4}&4&12\\ \hline
\frac{7}{4}&6&2\\ \hline
\frac{5}{4}&2&6\\ \hline
\frac{3}{4} &2&2\\ \hline
\frac{1}{4}&2&2\\ \hline
\end{array}
&
\eea
One can see that $\frac{5}{4}$ is the first energy level where an inequality of the degeneracies is produced
\bea
d_{\beta=\frac{3}{4}}(\frac{5}{4})&\neq &
d_{\beta=\frac{5}{4}}(\frac{5}{4}).\eea

The $\beta$-dependence of the vacuum energy is graphically presented in {\bf Figure 1}.

\begin{figure}
\begin{centering} 
 \includegraphics[width=0.60\textwidth, angle=0]{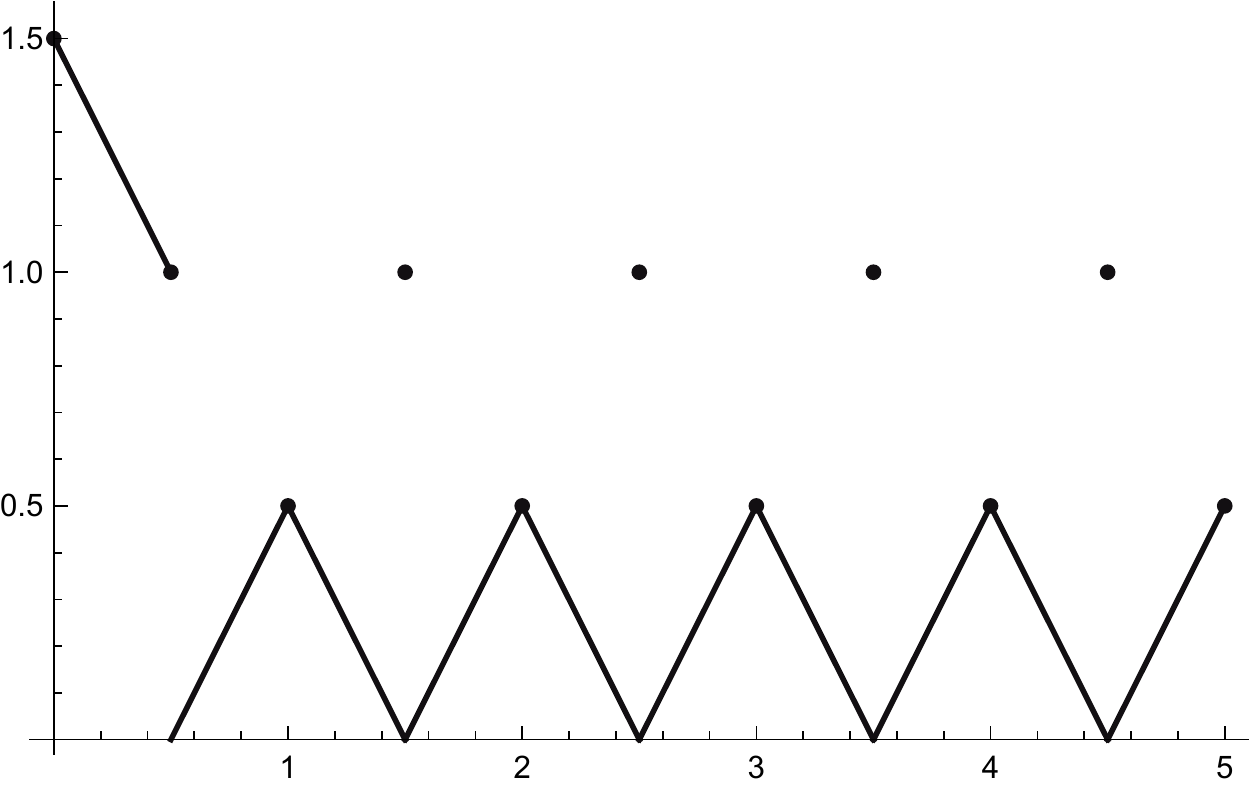}\\
%\begin{center}
{\bf Figure 1:}  the vacuum energy $E_{vac}(\beta)$ of the model is portrayed in the $y$ axis, with $\beta$ up to $\beta\leq 5$ depicted in the $x$ axis. This diagram refers to the Hilbert space admitting singular, but normalized wave functions at the origin. Starting from $\beta>\frac{1}{2}$, the graph is composed by a triangle wave of 
half-open line segments plus isolated points at $\beta=\frac{1}{2}+{\mathbb N}$. From $\beta\geq \frac{1}{2}$, the models with $\beta$ separated by an integer number and therefore with identical $\mu=\{ \beta-\frac{1}{2}\}$  possess the same spectrum, but different
degeneracies of their eigenvalues. For $\mu\neq0, \frac{1}{2}$, the same spectrum is also recovered (with different degeneracies) under the $\mu\leftrightarrow 1-\mu$ duality. The vacuum energy $E_{vac}(\beta)$ is always positive ($E_{vac}(\beta)>0$).
%\end{center}
\end{centering}
\end{figure}
%\newpage

\subsection{The spectrum for the ${\cal H}_{reg}$ Hilbert space of regular wave functions}

The $\beta=0$ case was already discussed since, for this value, the ${\cal H}_{norm}$ and ${\cal H}_{reg}$ Hilbert spaces coincide. For $\beta>0$ the following cases have to be separately treated:
\begin{itemize}
\item {\bf case ~~~ i}: ~ $0<\beta<\frac{1}{2}$,
\item {\bf case ~~ ii}: ~ $\beta= \frac{1}{2}$,
\item {\bf case ~ iii}: ~ $\frac{1}{2}<\beta<1$,
\item {\bf case ~~ iv}: ~ $\beta=1$,
\item {\bf case ~~~ v}: ~ $1<\beta< 2$, with $\beta\neq \frac{3}{2}$,
\item {\bf case ~~ vi}: ~ $\beta=\frac{3}{2}$,
\item {\bf case ~ vii}: ~ $\beta={2}$,
\item {\bf case ~viii}: ~ $\beta=2+\mu+p$, with $0<\mu<\frac{1}{2}$ and $p\in{\mathbb N}_0$,
\item {\bf case ~~ ix}: ~ $\beta=\frac{5}{2}+p$, with $p\in{\mathbb N}_0$,
\item {\bf case ~~~x}: ~ $\beta=2+\mu+p$, with $\frac{1}{2}<\mu<1$ and $p\in{\mathbb N}_0$,
\item {\bf case ~~ xi}: ~ $\beta=3+p$, with $p\in{\mathbb N}_0$.\bea
&&
\eea
\end{itemize}

The energy eigenvalues corresponding to the above cases are
\begin{itemize}
\item {\bf case ~ i}: ~ two series of energy eigenvalues, $E^+_n=\frac{3}{2}+\beta+n$ and $E^-_n=\frac{5}{2}-\beta+n$, with $n\in {\mathbb N}_0$, are encountered. \\
The vacuum energy is $E_{vac}=\frac{3}{2}+\beta$; the ground state is bosonic and doubly degenerate (hence, denoted as ``$2_B$").\\
The vacuum lowest weight vectors are specified by the quantum numbers $j=\frac{1}{2}$, $\delta=\epsilon=+1$.
\item {\bf case ~ ii}:  ~ the energy eigenvalues are $E_n=2+n$, with $n\in {\mathbb N}_0$.\\
The vacuum energy is $E_{vac}=2$; the degeneration of the ground state is $6$, with $2$ bosonic and $4$ fermionic eigenstates (``$2_B+4_F$").\\
The vacuum lowest weight vectors are specified by $j=\frac{1}{2}$, with $\delta=\epsilon=+1$ and by 
$j=\frac{3}{2}$ with $\delta=+1$ and $\epsilon=-1$.
\item {\bf case iii}:  ~ two series of energy eigenvalues, $E^+_n=\frac{3}{2}+\beta+n$ and $E^-_n=\frac{5}{2}-\beta+n$, with $n\in {\mathbb N}_0$, are encountered. \\
The vacuum energy is $E_{vac}=\frac{5}{2}-\beta$; the ground state is fermionic and $4$ times degenerate (``$4_F$").\\
The vacuum lowest vectors are specified by $j=\frac{3}{2}$, $\delta= +1$, $\epsilon =-1$.
\item {\bf case iv}: ~ the energy eigenvalues are $E_n=\frac{5}{2}+n$, with $n\in {\mathbb N}_0$.\\
The vacuum energy is $E_{vac}=\frac{5}{2}$; the degeneration of the ground state is $8$, with $2$ bosonic and $6$ fermionic eigenstates (``$2_B+6_F$").\\
The vacuum lowest weight vectors are specified by $j=\frac{1}{2}$, with $\delta=\epsilon=+1$ and by 
$j=\frac{5}{2}$ with $\delta=+1$ and $\epsilon=-1$.
\item {\bf case ~v}: ~  two series of energy eigenvalues $E_n^+= \frac{3}{2}+\beta+n$, $E_n^-=\frac{7}{2}-\beta+n$, with $n\in {\mathbb N}_0$, are encountered.\\
The vacuum energy is $E_{vac}=\frac{7}{2}-\beta$; the ground state is fermionic and $6$ times degenerate (``$6_F$"). \\
The vacuum lowest weight vectors are specified by $j=\frac{5}{2}$, $\delta=+1$, $\epsilon=- 1$.
\item {\bf case vi}: ~ the energy eigenvalues are $E_n=2+n$, with $n\in {\mathbb N}_0$.\\
The vacuum energy is $E_{vac}=2$; the degeneration of the ground state is fermionic and $6$ times degenerate (``$6_F$").\\
The vacuum lowest weight vectors are specified by
$j=\frac{5}{2}$ with $\delta=+1$ and $\epsilon=-1$.
\item {\bf case vii}:  ~ the energy eigenvalues are $E_n=\frac{5}{2}+n$, with $n\in {\mathbb N}_0$.\\
The vacuum energy is $E_{vac}=\frac{5}{2}$; the ground state is fermionic and $8$ times degenerate (``$8_F$").\\
The vacuum lowest weight vectors are specified by $j=\frac{7}{2}$, with $\delta=+1$ and $\epsilon=-1$.
\item {\bf case viii}: ~  two series of energy eigenvalues $E_n^-= \frac{3}{2}+\mu+n$, $E_n^+=\frac{5}{2}-\mu+n$, with $n\in {\mathbb N}_0$, are encountered.\\
The vacuum energy is $\frac{3}{2}+\mu$; the ground state is bosonic and $2(p+1)$ degenerate (``$(2p+2)_B$").\\
The vacuum lowest vectors are specified by $j=\frac{1}{2}+p$, with $\delta =-1$ and $\epsilon =+1$.
\item {\bf case ix}: ~ the energy eigenvalues are $E_n=2+n$, with $n\in {\mathbb N}_0$.\\
The vacuum energy is $E_{vac}=2$; the degeneration of the ground state is $10+4p$ with $2(p+1)$ bosonic and
$8+2p$ fermionic eigenstates (``$(2p+2)_B+(8+2p)_F$").\\
The vacuum lowest weight vectors are specified by $j=\frac{1}{2}+p$ with $\delta=-1$ and $\epsilon=+1$ and by $j=\frac{7}{2}+p$, with $\delta=+1$ and $\epsilon=-1$.
\item {\bf case x}: ~  two series of energy eigenvalues $E_n^+= \frac{3}{2}+\mu+n$, $E_n^-=\frac{5}{2}-\mu+n$, with $n\in {\mathbb N}_0$, are encountered.\\
The vacuum energy is $\frac{5}{2}-\mu$; the ground state is fermionic and $2(p+4)$ degenerate (``$(2p+8)_B$").\\
The vacuum lowest vectors are specified by $j=\frac{7}{2}+p$, with $\delta =+1$ and $\epsilon =-1$.
\item {\bf case xi}:  ~ the energy eigenvalues are $E_n=\frac{5}{2}+n$, with $n\in {\mathbb N}_0$.\\
The vacuum energy is $E_{vac}=\frac{5}{2}$; the degeneration of the ground state is $12+4p$ with $2(p+1)$ bosonic and
$10+2p$ fermionic eigenstates (``$(2p+2)_B+(2p+10)_F$").\\
The vacuum lowest weight vectors are specified by $j=\frac{1}{2}+p$ with $\delta=-1$ and $\epsilon=+1$ and by $j=\frac{9}{2}+p$, with $\delta=+1$ and $\epsilon=-1$.
\bea
&&\label{regcases}
\eea
\end{itemize}

The $\beta$-dependence of the vacuum energy is graphically presented in {\bf Figure 2}.
%\newpage
\begin{figure}
\begin{centering} 
 \includegraphics[width=0.60\textwidth, angle=0]{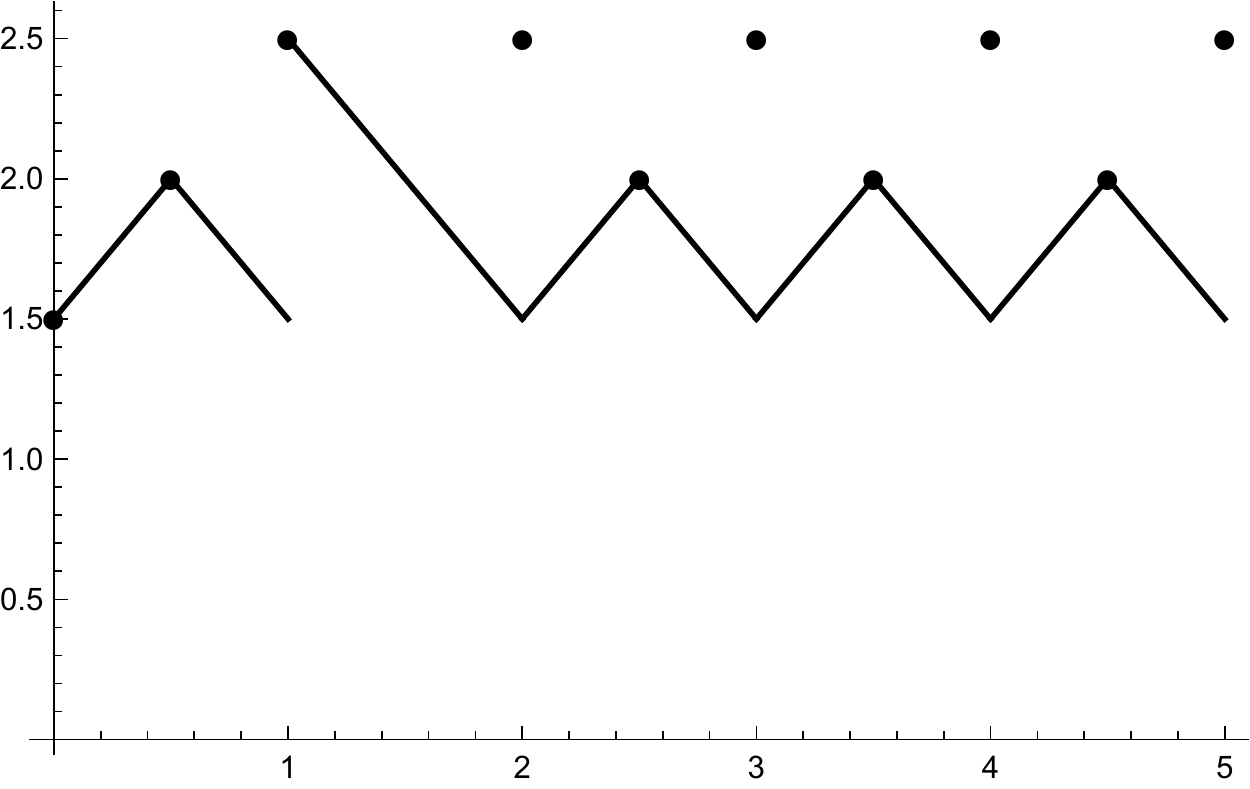}\\
%\begin{center}
{\bf Figure 2:}  the vacuum energy $E_{vac}(\beta)$ of the model is portrayed in the $y$ axis, with $\beta$ up to $\beta\leq 5$ depicted in the $x$ axis. This diagram refers to the Hilbert space satisfying the condition that its wave functions are regular at the origin. For $\beta>0$, the vacuum energy is always comprised in the interval $\frac{3}{2}<E_{vac}(\beta)\leq\frac{5}{2}$. The isolated points are now encountered at integer values of $\beta$ for $\beta=2,3,4,\ldots$. The features of this diagram are discussed in the main text.
%\end{center}
\end{centering}
\end{figure}

\subsection{About superselecting the wave functions}

It is beyond the scope of this work to present an investigation of the admissible superselection rules which can be implemented on the ${\cal H}_{norm}$ and ${\cal H}_{reg}$ Hilbert spaces. We limit ourselves to mention here
a natural type of superselection for its consequence on the spectrum. As we have seen, when the deformation parameter $\beta$ is integer or half-integer, the spectrum coincides with a shifted version of the undeformed oscillator (more on that in Section {\bf 6}).  A natural superselection consists in imposing the ground state and
the even excited states to be bosonic, with fermionic odd excited states (or viceversa, exchanging the role of bosons and fermions).  For $\beta\in {\mathbb Z}$ or $\beta\in \frac{1}{2}+{\mathbb Z}$ this superselection is implemented
by the projectors
\bea\label{proj1}
P_\pm &=&\frac{1}{2}({\mathbb I}_4 \pm N_Fe^{\pi i (H_{osc}- E_{gr})}) ,  \quad\quad (P_\pm^2=P_\pm),
\eea
where $E_{gr}$ denotes the given ground energy.
The ``$+$" (``$-$") sign is associated with the fermionic (bosonic) ground state.  The superselected wave functions $\Psi$ are
constrained to satisfy 
\bea
\textrm{either}\quad\quad P_+\Psi = 0, &\quad \textrm{or}\quad\quad P_-\Psi=0.
\eea 
\section{Combinatorics of the energy eigenstates degeneracy}

For $\beta$ integer or half-integer the spectrum of the deformed oscillator is a shifted version of the $\beta=0$ undeformed
oscillator. A second dissimilarity consists in the different degeneracies of the energy eigenvalues for the corresponding ground states and $n^{th}$ excited states. When $\beta {\not \in} \{{\mathbb Z}\}\cup\{\frac{1}{2}+{\mathbb Z}\}$, the spectrum is a combination of
two differently shifted spectra of the undeformed oscillator (see cases {\bf IV}, {\bf V}, {\bf VI} in (\ref{normcases}) and
cases {\bf i}, {\bf iii}, {\bf v}, {\bf viii}, {\bf x} in (\ref{regcases})).\par
In order to facilitate the comparison with the undeformed oscillator and highlight the differences, for $\beta$ integer and half-integer  we solve the combinatorics
of the degenerate eigenstates provided by the (\ref{iterative}) recursive relation with the (\ref{jadmissible}) input of primary (i.e. lowest weight vectors) states. The same technique can be applied for any real value of $\beta$ (the results for this latter case will not be reported to avoid eccessively burdening the paper).\par
Let $E_{\beta, norm}^n$ and $E_{\beta, reg}^n$ denote the energy eigenvalues  for $\beta$ integer or half-integer in correspondence with, respectively, the ${\cal H}_{norm}$ and ${\cal H}_{reg}$ choices of the Hilbert space ($E_{\beta  , norm}^0$, $E_{\beta  , reg}^0$ are the corresponding vacuum energies). The energy shifts $\Delta_{\beta, norm}$, $\Delta_{\beta, reg}$ with respect to the undeformed oscillator can be read from {\bf Figures 1} and {\bf 2}. We have
\bea
E_{\beta, norm}^n =E_{\beta=0}^n +\Delta_{\beta, norm}= \frac{3}{2}+n+ \Delta_{\beta, norm},&&
E_{\beta, reg}^n =E_{\beta=0}^n +\Delta_{\beta, reg}= \frac{3}{2}+n+ \Delta_{\beta, reg},\nonumber\\
\eea
with, for $p\in {\mathbb N}_0$,
\bea
&\Delta_{\frac{1}{2}+p, norm}=-\frac{1}{2},\quad\quad
\Delta_{p, norm}=-1,\quad\quad
\Delta_{\frac{1}{2}+p, reg}=\frac{1}{2},\quad\quad
\Delta_{p, reg}=1.&
\eea
We start our analysis of  the eigenvalue degeneracies by recalling, at first, the features of the $\beta=0$ undeformed
oscillator.

\subsection{Degeneracies of the $\beta=0$ undeformed oscillator}

Since, at $\beta=0$, the (\ref{hamosc}) Hamiltonian $H_{osc}$ corresponds to four copies of the ordinary isotropic three-dimensional
oscillator, its degeneracy $d_{\beta=0}(n)$ is
\bea
d_{\beta=0}(n)= 4\cdot d(n),  &\quad& \textrm{with}\quad \quad
d(n) = \frac{1}{2}(n^2+3n+2).
\eea
Here $d(n)$ denotes the eigenvalues degeneracy of the ordinary three-dimensional oscillator. It produces  the $1,3,6,10,15,\ldots $ tower of states, the vacuum being non-degenerate.\par
The rotational invariance of the ordinary three-dimensional oscillator implies that at each energy level the degenerate states can be accommodated into
multiplets (denoted as ``$[l]$") of non-negative integer orbital angular momentum $l$, each $[l]$ multiplet containing $2l+1$ states. We get, at the lowest orders,
\bea
&\begin{array}{ccll}
n=0 : &[0], &\textrm{so that} \quad d(0) = 1,\\
n=1 : &[1], & \textrm{so that} \quad d(1) = 3,\\
n=2 : &[0]\oplus [2], & \textrm{so that} \quad d(2) = 1+5=6,\\
n=3 : & [1]\oplus [3], & \textrm{so that} \quad d(3) = 3+7=10,\\
n=4 :  &[0]\oplus [2]\oplus [4], &\textrm{so that} \quad d(4) =1+5+9=15,\\
{\ldots}\quad ~: &{\ldots}, &\quad\quad\quad{\ldots}.
\end{array}&
\eea
The spectrum of the ordinary oscillator can be recovered by imposing, to the $H_{osc}$ Hamiltonian at $\beta=0$, two
independent superselection rules defined by the Fermion Parity Operator $N_F$ and by the spin operator $ {\mathbb I}_2\otimes{S_3}$ (since the spin-orbit coupling is absent at $\beta=0$, for this particular value of $\beta$ the Hermitian
operator ${\mathbb I}_2\otimes S_3$ commutes with the $H_{osc}$ Hamiltonian). The corresponding projector
operators $P_\pm^N$, $P_\pm^S$   which induce the superselection rules are different from the projectors introduced in (\ref{proj1}); they are given by 
\bea
P_\pm^N = \frac{1}{2}({\mathbb I}_4\pm N_F), &\quad& 
P_\pm^S = \frac{1}{2}({\mathbb I}_4\pm 2 {\mathbb I}_2\otimes S_3)
\eea
and satisfy
\bea
&(P_\pm^N)^2= P_\pm^N, \quad\quad(P_\pm^S)^2 =P_\pm^S, \quad\quad [P_\pm^N, P_\pm^S]=0,\quad\quad [P_\pm^N, P_\mp^S]=0.&
\eea
The two compatible superselection rules at $\beta=0$ can be imposed by restricting the $H_{osc}$ wave functions $\Psi$ to satisfy
\bea
&P_+^N\Psi=P_+^S\Psi=\Psi.&
\eea

One should note that the $\beta=0$ rotational invariance is, from the $sl(2|1)$ spectrum-generating superalgebra viewpoint,
an emergent symmetry, being not immediately derivable from the $sl(2|1)$ data.

\subsection{Degeneracies for $\beta=\frac{1}{2} +{\mathbb N}_0$ and $\beta=1 +{\mathbb N}_0$ with ${\cal H}_{norm}$ Hilbert space}

We present here the results of the degeneracies for the Hilbert space ${\cal H}_{norm}$ of normalized wave functions. We have
\par
{\bf Case a}: $\beta=\frac{1}{2} +p$ (energy levels $E_n = n+1$)  with $p,n\in {\mathbb N}_0$.\par

For $\beta=\frac{1}{2}$ the degeneracy $d_{\beta=\frac{1}{2}}(E_n)$ of the $n^{th}$ level is simply
\bea
d_{\beta=\frac{1}{2}}(E_n)&=& 2\cdot (n+1)^2.
\eea
The above formula can be recovered from the following most general case at given integer $p$.  The degeneracy 
$d_{\beta=\frac{1}{2}+p}(E_n)$ grows linearly
(mimicking a two-dimensional oscillator)
up to $n=p$; it then grows quadratically starting from $n= p+1$:
\bea
d_{\beta=\frac{1}{2}+p}(E_n)&=& 2(n+1)(2p+1) \quad \textrm{for} \quad n=0,1,2,\ldots, p,\nonumber\\
d_{\beta=\frac{1}{2}+p}(E_n)&=& 2\cdot(q^2+2(p+1)q+(p+1)(2p+1))\quad \textrm{for} \quad n=p+q\quad \textrm{with}\quad q=0,1,2,\ldots.\nonumber\\
&&
\eea
{\bf Case b}: $\beta=1 +p$ (energy levels $E_n = n+\frac{1}{2}$)  with $p,n\in {\mathbb N}_0$.\par
As in the previous case, the degeneracy 
$d_{\beta=1+p}(E_n)$ grows linearly
(mimicking a two-dimensional oscillator)
up to $n=p$; it then grows quadratically starting from $n= p+1$:
\bea
d_{\beta=1+p}(E_n)&=& 4(n+1)(p+1) \quad \textrm{for} \quad n=0,1,2,\ldots, p,\nonumber\\
d_{\beta=1+p}(E_n)&=& 2\cdot(q^2+(2p+1)q+2(p+1)^2)\quad \textrm{for} \quad n=p+q\quad \textrm{with}\quad q=0,1,2,\ldots.\nonumber\\
&&
\eea
We provide in {\bf Figure 3} a graphical description of the degeneracy.
%\newpage
\begin{figure}
\begin{centering}
 \includegraphics[width=0.70\textwidth, angle=0]{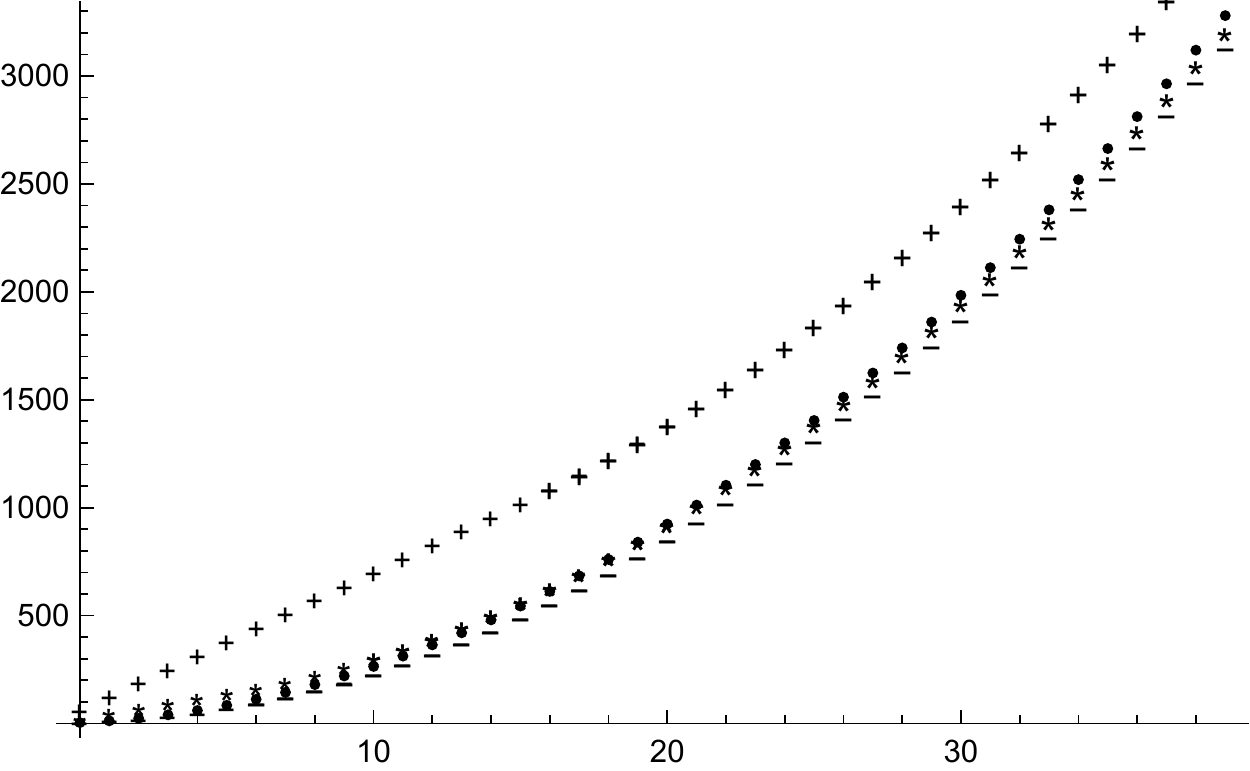}
%\begin{center}

{\bf Figure 3:}  the energy degeneracy ($y$ axis) for the ${\cal H}_{norm}$ Hilbert space at the integer values \\
\quad $\beta=0,2,6,16$. In the $x$ axis are reported the $40$ lowest energy eigenvalues. The ``$\bullet$" bullet\\ denotes the $\beta=0$ undeformed oscillator, while  ``$-$", ``$\ast$" and ``$+$" stand, respectively, for the $\beta=2, 6, 16$, cases. One can note the ``bending" of the $\beta=16$ curve around energy $E=16$.
%\end{center}
%\end{centering}
%\%end{centering}
%\end{figure}

\end{centering}
\end{figure}

\subsection{Degeneracies for $\beta=\frac{1}{2} +{\mathbb N}_0$ and $\beta=1 +{\mathbb N}_0$ with ${\cal H}_{reg}$ Hilbert space}

We present here the results of the degeneracies for the Hilbert space ${\cal H}_{reg}$ of regular wave functions. We have
\par
{\bf Case a}: $\beta$ half-integer with energy levels $E_n = n+2$. \par

The $\beta=\frac{1}{2}$ and $\beta=\frac{3}{2}$ cases present the same degeneracy, the functions
$d_{\beta=\frac{1}{2}}(E_n)$, $d_{\beta=\frac{3}{2}}(E_n)$ being given by
\bea
d_{\beta=\frac{1}{2}}(E_n)=d_{\beta=\frac{3}{2}}(E_n) &=& 2\cdot(n^2+4n+3).
\eea
The $\beta=\frac{3}{2}$ case can also be recovered from the most general formula for $\beta=\frac{3}{2}+p$, with $p\in {\mathbb N}_0$. We have
\bea
d_{\beta=\frac{3}{2}+p}(E_n)&=&2 \cdot (n+1)(2p+3) \quad \textrm{for} \quad n=0,1,2,\ldots, p,\nonumber\\
d_{\beta=\frac{3}{2}+p}(E_n)&=& 2\cdot(q^2+(2p+4)q+(2p+3)(p+1))\quad \textrm{for} \quad n=p+q\quad \textrm{with}\quad q=0,1,2,\ldots.\nonumber\\
&&
\eea
We recover the same feature observed for normalized wave functions belonging to ${\cal H}_{norm}$, namely a linear growth followed by a quadratic growth.\par
{\bf Case b}: $\beta$ integer with energy levels $E_n = n+\frac{5}{2}$. \par

The $\beta=1$ and $\beta=2$ cases present the same degeneracy, the functions $d_{\beta=1}(E_n)$, $d_{\beta=2}(E_n)$ being given by
\bea
d_{\beta=1}(E_n)=d_{\beta=2}(E_n) &=& 2\cdot(n^2+5n+4).
\eea

The $\beta=2$ case can also be recovered from the most general formula for $\beta=2+p$, with $p\in {\mathbb N}_0$. We have
\bea
d_{\beta=2+p}(E_n)&=&4 \cdot (p+2)(n+1) \quad \textrm{for} \quad n=0,1,2,\ldots, p,\nonumber\\
d_{\beta=2+p}(E_n)&=& 2\cdot(q^2+(2p+5)q+2(p+2)(p+1))\quad \textrm{for} \quad n=p+q\quad \textrm{with}\quad q=0,1,2,\ldots.\nonumber\\
&&
\eea
Here again we observe a linear growth up to $E\approx \beta$, followed by a quadratic growth.\par

\section{Dimensional reductions of the $3D$ deformed oscillator}

In this Section we discuss how to recover two-dimensional and one-dimensional models of superconformal quantum
mechanics (originally introduced in \cite{cuhoto} via the quantization of worldline superconformal sigma-models, with two and one target coordinates, respectively) by dimensional reduction of the three-dimensional superconformal quantum mechanics. We explicitly treat the case of the de Alfaro-Fubini-Furlan deformed oscillator (the construction can be
repeated step by step, with analogous results, in the absence of the oscillatorial term).

\subsection{The $3D\rightarrow 2D$ dimensional reduction}

To obtain this dimensional reduction we have to freeze the dependence on the third coordinate $x_3$ in the wave functions. This can be achieved by setting $\slashed{\partial}$, $\slashed{r}$, $r$ entering the operators
(\ref{qaop},\ref{hop},\ref{kop},\ref{remop}) to be restricted to
\bea\label{restrictd2}
&\slashed{\partial}=h_1\partial_1+h_2\partial_2,\quad\quad {\slashed{r}}= x_1h_1+x_2h_2,\quad\quad r=\sqrt{x_1^2+x_2^2}&
\eea
(the last equality holds for $D$  in formula (\ref{remop})). With the above positions the $sl(2|1)$ (anti)commutators
(\ref{sl21comanticom}) are maintained. The ${\overrightarrow{{\mathbf{S}}}}\cdot{\overrightarrow{\mathbf{L}}}$
operator entering the Hamiltonians (\ref{hop}) and (\ref{hamosc}) is now given by
$S_3L_3$. In the two-dimensional case, being proportional to $\sigma_3$, this operator is diagonal.\par
The resulting Hamiltonian $H_{2D,osc}$ of the deformed oscillator corresponds to two copies of the two-dimensional $2\times 2$ matrix Hamiltonians  derived in \cite{cuhoto}
from the quantization of the $sl(2|1)$ worldline sigma-model with two propagating bosonic and two propagating fermionic fields. We have
\bea\label{ham2d}
H_{2D,osc}&=&-\frac{1}{2}(\partial_{x_1}^2+\partial_{x_2}^2)\cdot {\mathbb I}_4+\frac{1}{2r^2}(\beta^2{\mathbb I}_4+
\beta N_F(1+2\cdot {\mathbb I}_2\otimes \sigma_3L_3))+\frac{1}{2}r^2{\mathbb I}_4.
\eea
The two non-vanishing upper-left  and  bottom-right blocks  of the given $4\times 4$ Hamiltonian present the deformation
parameters $+\beta$ and $-\beta$, respectively.\par
There is an important remark. For the deformed oscillator, the $sl(2|1)$ spectrum-generating superalgebra  induced by the (\ref{qaop},\ref{hop},\ref{kop},\ref{remop}) operators with the (\ref{restrictd2}) restriction does not
coincide with the $sl(2|1)$ spectrum-generating superalgebra introduced in \cite{cuhoto}. The reason is that the
ladder operators constructed with $Q_a$ and ${\overline Q}_a$ connect the upper-left and bottom-right blocks,
while the $sl(2|1)$ ladder operators of \cite{cuhoto} act inside each $2\times 2$ component blocks. Therefore, the
$sl(2|1)$ spectrum-generating superalgebra of \cite{cuhoto}, as well as its charge-conjugated $sl(2|1)$ superalgebra (also in \cite{cuhoto}) are, from the point of view of the dimensionally reduced theory, emergent spectrum-generating
superalgebras which are not directly connected with the original $sl(2|1)$ superalgebra.

\subsection{The $3D\rightarrow 1D$ dimensional reduction}

To operate this dimensional reduction it is convenient to freeze the dependence on the $x_1$, $x_2$ coordinates by
setting
\bea\label{restrictd1}
&\slashed{\partial}=h_3\partial_3,\quad\quad {\slashed{r}}= x_3h_3,\quad\quad r=\sqrt{x_3^2}.&
\eea
The (\ref{qaop},\ref{hop},\ref{kop},\ref{remop}) operators continue, with the above positions, to satisfy the $sl(2|1)$ (anti)commutators
(\ref{sl21comanticom}).
The resulting $H_{1D,osc}$ deformed oscillator, given by (we set, for simplicity, $x=x_3$)
\bea
H_{1D,osc}&=&-\frac{1}{2}\partial_x^2\cdot{\mathbb I}_4 +\frac{1}{2x^2}(\beta^2\cdot {\mathbb I}_4+\beta N_F)+\frac{1}{2}x^2\cdot {\mathbb I}_4,
\eea
coincides with the model derived in \cite{cuhoto} from the quantization of the world-line sigma model induced by the $(1,4,3)$ supermultiplet (namely, with one propagating boson, four propagating fermions and three auxiliary fields). 
The $H_{1D,osc}$ Hamiltonian possesses the larger $D(2,1;\alpha)$ spectrum-generating superalgebra, with $\alpha=\beta-\frac{1}{2}$. On the other hand, as recalled in \cite{ackt}, the $sl(2|1)\subset D(2,1;\alpha)$ generators are sufficient
to determine the ray vectors of the Hilbert space (the vectors produced by the remaining generators differ by an inessential phase). From the dimensional reduction viewpoint, the extra generators entering $D(2,1;\alpha)$ are associated with an emergent symmetry, not manifest from the $sl(2|1)$ construction.

\section{Conclusions}

We provided a direct construction of a three-dimensional superconformal quantum mechanics and of its associated
de Alfaro-Fubini-Furlan deformed oscillator. Our approach allowed to overcome the difficulties (pinpointed in \cite{cuhoto}) in quantizing worldline superconformal sigma-models with $D>2$ target dimensions. It is rewarding that
a dimensional reduction of the model allows to recover the $D=1,2$ quantized worldline superconformal sigma-models introduced in \cite{cuhoto}. For the $3D$ deformed oscillator, $sl(2|1)$ acts as a spectrum-generating superalgebra;  the complete spectrum is obtained from a (infinite) direct sum of $sl(2|1)$ lowest weight representations. Depending on the coupling constant $\beta$, different Hilbert spaces can be consistently selected
by suitably restricting the allowed $sl(2|1)$ lowest weight representations.\par
Even if $sl(2|1)$ is sufficient to completely determine the spectrum, extra symmetries can play a role
(possibly as enhanced symmetries at given values of the coupling constant $\beta$). This is worth investigating since
the recalled arbitrariness in the selection of the Hilbert space can be eliminated if an enlarged algebra is responsible
for extra conditions to be imposed.\par
In this paper we were able to determine the three-dimensional quantum Hamiltonians (given in (\ref{hop}) and (\ref{hamosc}) for the superconformal and, respectively, deformed oscillator cases). Once the Hamiltonian has been individuated, there is a standard method to be applied (see \cite{olv}) to derive the symmetry operators of the associated partial differential equation (for the case at hand, the partial differential equation is the time-dependent
Schr\"odinger equation for the given Hamiltonian). By applying this method, the most general dynamical symmetry
of the model can be derived through a very lengthy, but straightforward procedure. We plan to address this investigation in a future work. It allows to answer the open question about the possible existence of extra symmetries
and of the role they could play. It is worth mentioning that a similar analysis, conducted for a one-dimensional superconformal quantum mechanics, led to the surprising result \cite{tv} that the $AdS_2$ higher-spin superalgebra \cite{vas} is a dynamical symmetry of the model.
A tantalizing possibility is that the extra symmetries, as those mentioned in Appendix {\bf C},  of the superconformal quantum mechanics could
be linked to a BMS/CFT correspondence (see \cite{bt}), the BMS transformations being associated with the ``large" diffeomorphisms acting on asymptotically flat space-times \cite{str}.
\par
~\\
{~}
%\par
%\newpage
\renewcommand{\theequation}{A.\arabic{equation}}
\setcounter{equation}{0}
 {~}\par
{\bf{\Large{Appendix A: notations and conventions}}}
~\par
~\par
We collect here for convenience the relevant notations and conventions used throughout the paper.\par
The $n\times n$ Identity matrix is denoted as ``${\mathbb I}_n$". The three Pauli matrices $\sigma_i$ are
\bea\label{pauli}
&{\footnotesize{\sigma_1=\left(\begin{array}{cc} 0&1\\
1&0\end{array}\right),\quad \sigma_2=\left(\begin{array}{cc} 0&-i\\
i&0\end{array}\right),\quad \sigma_3=\left(\begin{array}{cc} 1&0\\
0&-1\end{array}\right).}}&
\eea
The following $4\times 4$ complex matrices $\gamma_j, h_j$ ($j=1,2,3$) are introduced:
\bea\label{4x4matrices}
\gamma_j = \sigma_j\otimes {\mathbb I}_2, \quad &&\quad h_j = i{\mathbb I}_2\otimes \sigma_j.
\eea
The matrices $\gamma_1,\gamma_2$ are block-antidiagonal, while $\gamma_3=-i\gamma_1\gamma_2$ is the
Fermion Parity Operator $N_F$ which defines bosons and fermions as its eigenvectors with eigenvalues $+1$  and $-1$, respectively.  When we need to stress this property, we use the notation
\bea
N_F=\gamma_3.\eea
The imaginary unit ``$i$" is introduced in the definition of the matrices $h_j$, so that they furnish
a representation of the three imaginary quaternions. Indeed, the composition law
\bea\label{quaternions}
h_i\cdot h_j &=& -\delta_{ij}{\mathbb I}_4 -\epsilon_{ijk}h_k
\eea
is satisfied. 
The totally antisymmetric tensor $\epsilon_{ijk}$ is normalized so that $\epsilon_{123}=1$. Throughout the paper the Einstein convention of summation over the repeated indices is understood unless otherwise specified.\par
By construction, the matrices $\gamma_i$ commute with the quaternionic matrices $h_j$:
\bea
\relax
[\gamma_i,h_j]&=&0\quad \forall~ i,j=1,2,3.
\eea
In Cartesian coordinates the position vector ${\overrightarrow{\bf r}}$ has components $x_i$. The orbital angular momentum 
${\overrightarrow{\bf L}}$ has components $L_i$, where 
\bea
L_i &=& -i\epsilon_{ijk}x_j\partial_k.
\eea
For convenience, we introduced the slashed notation for the Euclidean version of the Dirac's operator and a 
matrix-valued space coordinates operator, by setting 
\bea\label{rhLh}
{\slashed{\partial}} := {\overrightarrow {\bf \nabla}}\cdot {\overrightarrow {\bf h}}= \partial_ih_i , & \quad&
{\slashed{r}} := {\overrightarrow {\bf r}}\cdot {\overrightarrow {\bf h}}= x_ih_i.
\eea
It follows, in particular, that ${\slashed{\partial}}^2=-\nabla^2$ and ${\slashed{r}}^2=-r^2$, where $
r=\sqrt{x_1^2+x_2^2+x_3^2}$ is the radial coordinate. In our conventions the spherical coordinates $r,\theta,\phi$
(restricted to $r\geq 0$, $0\leq \theta\leq \pi$, $0\leq \phi<2\pi$) are introduced through the positions
\bea
& x_1=r \sin\theta \cos\phi,\quad x_2=r \sin\theta \sin\phi,\quad x_3=r \cos\theta.
\eea
The spin ${\overrightarrow{\bf S}}$ has components $\frac{1}{2}\sigma_i$, so that
\bea\label{spin}
{\overrightarrow{\bf S}}=\frac{1}{2}{\overrightarrow{\sigma}}, \quad &{\textrm{then}}& \quad {\overrightarrow{\bf S}}^2=\frac{3}{4}{\mathbb I}_2 = s(s+1){\mathbb I}_2\quad {\textrm{with} }\quad s=\frac{1}{2}.
\eea
\par
~\\
{~}
%\par
%\newpage
\renewcommand{\theequation}{B.\arabic{equation}}
\setcounter{equation}{0}
 {~}\par
{\bf{\Large{Appendix B: the orthonormal eigenstates}}}
~\par
~\par
We present for completeness the orthonormal eigenstates. The derivation of the orthonormal conditions is based on some key observations and a lengthy, but straightforward inductive proof concerning
the excited states. We outline the main points and give the final results.\par
At first one has to mention that the spin spherical harmonics $\mathcal{Y}_{j,l,m}\left(\theta,\phi\right)$ introduced in (\ref{spinspherical}) are orthonormal, satisfying
\bea
\int d\Omega \mathcal{Y}^\dagger_{j,l,m}\left(\theta,\phi\right)\mathcal{Y}_{j',l',m'}\left(\theta,\phi\right)&=&\delta_{jj'}\delta_{ll'}\delta_{mm'},
\eea
with $d\Omega$ an infinitesimal solid angle.\par
By setting $y=r^2$, the integral (\ref{gammafromhere}) produces a Gamma function, so that
\bea
&\int_0^{+\infty}dr r^2 r^{2\gamma_{(j,\delta,\epsilon)}}e^{-r^2}=\frac{1}{2}\int_0^{+\infty}dy y^{\gamma_{(j,\delta,\epsilon)}+\frac{1}{2}}e^{-y}= \frac{1}{2}\Gamma(\gamma_{(j,\delta,\epsilon)}+\frac{3}{2}).&
\eea
The lowest weight vectors $\Psi_{j,\delta,m}^\epsilon(r,\theta,\phi)$ from (\ref{lwv}) which satisfy, at given $\beta$, the (\ref{lws}) condition are orthogonal. The orthonormal wave functions $\Psi_{N,j,\delta,m}^\epsilon(r,\theta,\phi)$ are expressed in terms of the normalizing factor $M^{\epsilon,\beta}_{j,\delta}$. We have
\bea\label{normalizing}
M^{\epsilon,\beta}_{j,\delta} = \sqrt{\frac{2}{\Gamma(\epsilon\beta+\delta j+\frac{1}{2}(1+\delta))}}, &\quad&
\Psi_{N,j,\delta,m}^\epsilon(r,\theta,\phi)=M^{\epsilon,\beta}_{j,\delta}\Psi_{j,\delta,m}^\epsilon(r,\theta,\phi),
\eea
so that
\bea
\langle \Psi_{N,j,\delta,m}^\epsilon(r,\theta,\phi)|\Psi_{N,j',\delta',m'}^{\epsilon'}(r,\theta,\phi)\rangle &=& \delta_{jj'}\delta_{mm'}\delta_{\epsilon\epsilon}\delta_{\delta\delta'}.
\eea
The excited eigenstates $({a_1^+})^k\Psi_{j,\delta,m}^\epsilon(r,\theta,\phi)$, obtained by applying $k$ times the $a_1^+$ creation operator (\ref{creation}), are orthogonal, as it can be easily verified. The computation of their normalization factors which make the wave functions orthonormal is more cumbersome. It involves the computation of Rodrigues-type formulas, see \cite{szego}, for recursive polynomials in the radial coordinate $r$. These recursive polynomials can be recovered from the associated Laguerre's polynomials. In order to proceed we recall that
\bea\label{a1+}
a_1^+ &=&\frac{1}{\sqrt 2}\gamma_1\frac{\slashed{r}}{r}({\mathbb I}_4\cdot(\partial_r- r)-\frac{2}{r}{\mathbb I}_2\otimes 
{\vec{\bf S}}\cdot{\vec{\bf L}}-\frac{\beta}{r}N_F)
\eea
and that the (unnormalized) lowest weight vectors $\Psi_{j,\delta,m}^\epsilon(r,\theta,\phi)$ from (\ref{lwv}) are 
\bea\label{lwvagain}
\Psi_{j,\delta,m}^\epsilon(r,\theta,\phi)&=& e_{\epsilon}\otimes \mathcal{Y}_{j,j-\frac{1}{2}\delta,m}\left(\theta,\phi\right)\cdot r^{\beta\epsilon+\delta j +\frac{1}{2}\delta-1}e^{-\frac{1}{2}r^2}.
\eea
The action of $ {\vec{\bf S}}\cdot{\vec{\bf L}}$ can be read from equation (\ref{spinorbitop}), while the action
of $\frac{\slashed{r}}{r}$ can be read from the
\bea
\frac{{\vec {\bf r}}\cdot {\vec{\bf{ \sigma}}}}{r}\mathcal{Y}_{j,j-\frac{1}{2}\delta,m}\left(\theta,\phi\right)&=&-\mathcal{Y}_{j,j+\frac{1}{2}\delta,m}\left(\theta,\phi\right)
\eea
identity. The even and odd excited states are therefore respectively given by
\bea\label{excitedstates}
(a_1^+)^{2k}\Psi_{j,\delta,m}^\epsilon(r,\theta,\phi)&=&e_{\epsilon}\otimes \mathcal{Y}_{j,j-\frac{1}{2}\delta,m}\left(\theta,\phi\right)\cdot (-2)^k p_{2k,j}^{\epsilon,\delta,\beta} (r) r^{\epsilon\beta+\delta j +\frac{1}{2}\delta-1}e^{-\frac{1}{2}r^2},
\nonumber\\
(a_1^+)^{2k+1}\Psi_{j,\delta,m}^\epsilon(r,\theta,\phi)&=&i\sqrt{2} e_{-\epsilon}\otimes \mathcal{Y}_{j,j+\frac{1}{2}\delta,m}\left(\theta,\phi\right)\cdot (-2)^k p_{2k+1,j}^{\epsilon,\delta,\beta} (r) r^{\epsilon\beta+\delta j +\frac{1}{2}\delta-1}e^{-\frac{1}{2}r^2},\nonumber\\&&
\eea
where $p_{2k,j}^{\epsilon,\delta,\beta} (r)$ and $p_{2k+1,j}^{\epsilon,\delta,\beta} (r)$ are $r$-dependent polynomials recursively determined by the Rodrigues-type formulas
\bea\label{poly}
p_{2k,j}^{\epsilon,\delta,\beta} (r)&=&\frac{1}{2^{2k}}\left(\begin{array}{cc}r^{-{\overline\gamma}}e^{\frac{r^2}{2}}&0\end{array}\right)\left(\begin{array}{cc}0&\partial_r-r+\frac{{\overline\gamma}+2}{r}\\\partial_r-r-\frac{{\overline\gamma}}{r}&0\end{array}\right)^{2k}\left(\begin{array}{c}r^{{\overline\gamma}}e^{-\frac{r^2}{2}}\\0\end{array}\right), \nonumber\\
p_{2k+1,j}^{\epsilon,\delta,\beta} (r) &=&\frac{1}{2^{2k+1}}\left(\begin{array}{cc}r^{-{\overline\gamma}}e^{\frac{r^2}{2}}&0\end{array}\right)\left(\begin{array}{cc}0&\partial_r-r+\frac{{\overline\gamma}+2}{r}\\\partial_r-r-\frac{{\overline\gamma}}{r}&0\end{array}\right)^{2k+1}\left(\begin{array}{c}0\\r^{{\overline\gamma}}e^{-\frac{r^2}{2}}\end{array}\right),\nonumber\\&&
\eea
where, see (\ref{gamma}), 
\bea\label{overlinegamma}
&{\overline\gamma} \equiv{\gamma_{(j,\delta,\epsilon)}}(\beta)=\epsilon\beta+\delta j+\frac{1}{2}\delta-1.&
\eea
The connection with the associated Laguerre's polynomials is as follows (we discuss it explicitly for the even polynomials 
$p_{2k,j}^{\epsilon,\delta,\beta} (r)$, the derivation for the odd polynomials $p_{2k+1,j}^{\epsilon,\delta,\beta} (r)$
is made along the same lines). We obtain, from (\ref{poly}), the relation
\bea
p_{2(k+1),j}^{\epsilon,\delta,\beta} (r) &=&\frac{1}{4}r^{-{\overline\gamma}}e^{\frac{r^2}{2}}(\partial_r-r+\frac{{\overline\gamma}+2}{r})(\partial_r-r-\frac{{\overline\gamma}}{r})
p_{2k,j}^{\epsilon,\delta,\beta} (r) r^{{\overline\gamma}}e^{-\frac{r^2}{2}}.
\eea
It  follows in particular, from $p_{0,j}^{\epsilon,\delta,\beta} (r)=1$, that
\bea
p_{2,j}^{\epsilon,\delta,\beta} (r) &=& r^2-{\overline\gamma}-\frac{3}{2}.
\eea
The associated Laguerre polynomials $L_k^{(\gamma)}(x)$ are introduced through the position
\bea
L_k^{(\gamma)}(x)&=& \frac{x^{-\gamma}e^x}{k!}(\frac{d}{dx})^k x^{\gamma+k}e^{-x}.
\eea
They satisfy the identities
\bea\label{lagidentities}
L_k^{(\gamma)}(x) &=& L_k^{(\gamma+1)}(x)-L_{k-1}^{(\gamma+1)}(x),\nonumber\\
xL_{k-1}^{(\gamma+1)}(x) &=& (\gamma+k)L_{k-1}^{(\gamma)}(x)-kL_k^{(\gamma)}(x).
\eea
Since
\bea
L_1^{(\gamma)}(x) &=& -x+\gamma-1,
\eea
by setting
\bea
x=r^2, &\quad&\gamma={\overline \gamma}+\frac{1}{2},
\eea
we can identify
\bea
p_{2,j}^{\epsilon,\delta,\beta} (r) &=&-L_1^{({\overline\gamma}+\frac{1}{2})}(r^2).
\eea
By assuming the Ansatz
\bea\label{ansatz}
p_{2k,j}^{\epsilon,\delta,\beta} (r) &=&C_kL_k^{({\overline\gamma}+\frac{1}{2})}(r^2),
\eea
after lengthy computations and the use of the (\ref{lagidentities}) identities, one arrives at the inductive
proof that (\ref{ansatz}) is satisfied, provided that 
\bea
C_k&=& (-1)^k k!
\eea
The $p_{2k,j}^{\epsilon,\delta,\beta} (r)$ even and $
p_{2k+1,j}^{\epsilon,\delta,\beta} (r)$ odd polynomials are expressed, in terms of the associated Laguerre
polynomials, as
\bea
p_{2k,j}^{\epsilon,\delta,\beta} (r)&=&(-1)^k k! L_k^{({\overline\gamma}+\frac{1}{2})}(r^2),\nonumber\\
p_{2k+1,j}^{\epsilon,\delta,\beta} (r)&=&(-1)^{k+1} k! rL_k^{({\overline\gamma}+\frac{3}{2})}(r^2).
\eea
By plugging these expressions into the (\ref{excitedstates}) formulas we are able to determine the normalizing factors that need to be used to construct orthonormal excited eigenstates. The normalizing factors are recovered from the orthogonal relations for the associated Laguerre polynomials, given by
\bea
\int_0^{+\infty}dx x^\gamma e^{-x}L_n^{(\gamma)}(x)L_m^{(\gamma)}(x)&=&\frac{\Gamma(n+\gamma+1)}{n!}\delta_{nm}.
\eea
The final expressions for the orthonormal wave functions 
$\Psi_{N, k,j,\delta,m}^\epsilon(r,\theta,\phi)$ are
\bea\label{eveneigen}
\Psi_{N, 2k,j,\delta,m}^\epsilon(r,\theta,\phi)&=&e_{\epsilon}\otimes \mathcal{Y}_{j,j-\frac{1}{2}\delta,m}\left(\theta,\phi\right)\cdot M_{2k}^{\overline \gamma}L_k^{({\overline\gamma}+\frac{1}{2})}(r^2)\cdot r^{\overline\gamma} e^{-\frac{r^2}{2}}
\eea
with
\bea
M_{2k}^{\overline \gamma}&=& \sqrt{\frac{(k!)\cdot 2}{\Gamma(k+{\overline\gamma}+\frac{3}{2})}}
\eea
and
\bea\label{oddeigen}
\Psi_{N, 2k+1,j,\delta,m}^\epsilon(r,\theta,\phi)&=&e_{-\epsilon}\otimes \mathcal{Y}_{j,j+\frac{1}{2}\delta,m}\left(\theta,\phi\right)\cdot M_{2k+1}^{\overline \gamma}L_k^{({\overline\gamma}+\frac{3}{2})}(r^2)\cdot r^{{\overline\gamma}+1} e^{-\frac{r^2}{2}}
\eea
with
\bea
M_{2k+1}^{\overline \gamma}&=& \sqrt{\frac{(k!)\cdot 2}{\Gamma(k+{\overline\gamma}+\frac{5}{2})}}.
\eea
The parameter ${\overline\gamma} ={\gamma_{(j,\delta,\epsilon)}}(\beta)$ was introduced in (\ref{overlinegamma}).
One should note that $M_0^{\overline\gamma}$ reproduces, as it should be, the normalizing factor given in formula (\ref{normalizing}).
~
\par
~\\
{~}
%\par
%\newpage
\renewcommand{\theequation}{C.\arabic{equation}}
\setcounter{equation}{0}
 {~}\par
{\bf{\Large{Appendix C: a graphical illustration of an open problem}}}
~\par
~\par
The three dimensional deformed oscillator can be completely solved, its energy spectrum and associated degeneracy computed. On the other hand, as discussed in Sections {\bf 5} and {\bf 6}, alternative admissible choices of Hilbert space lead to different results corresponding to different quantum models. This means that the spectrum-generating superalgebra provides a (relevant) piece of information, but
it does not allow to uniquely deduce the spectrum and the degeneracy of the $\beta$-deformed model. Some extra information should be supplemented. It is quite
possible that (yet to be understood) larger algebraic structures could be responsible for that, in such a way that they allow to uniquely pinpoint a given quantum model. 
The source of this ambiguity is the fact that the Hilbert space is not expressed by a single irreducible representation, but by a direct sum of several (infinite) lowest weight representations of the $sl(2|1)$ spectrum-generating superalgebra.  Consistent selections of subsets of the set of lowest weight representations therefore lead to different Hilbert spaces.  \par
The open problem of searching for enlarged algebraic structures is not specific to the three-dimensional deformed oscillator;  it applies to the whole class of related (deformed) oscillators. It already appears in the case
of the two-dimensional oscillator which corresponds, see formula (\ref{ham2d}), to a dimensional reduction of the $3D$ oscillator. It is quite appealing to focus on this simpler case because it offers a nice graphical visualization of the question at hand.\par
For our illustrative purpose it is sufficient to take a superselected version of the undeformed two-dimensional theory with a unique bosonic vacuum and integerly-spaced energy excited states (the energy degeneracy growing linearly) arranged in  a triangular shape as shown in Figures {\bf 4} and {\bf 5} below. As discussed in Section {\bf 3}, the ray vectors are determined by the $osp(1|2)\subset sl(2|1)$ spectrum generating subalgebra (the action of the remaining $sl(2|1)$ generators produce inessential phases). In \cite{cuhoto}
it was shown that a new set of operators, obtained by conjugating the original operators of the $sl(2|1)$ 
spectrum-generating superalgebra,  produce another $sl(2|1)$ superalgebra, denoted as $sl(2|1)_C$.  The details of the \cite{cuhoto} construction are not relevant here. What is relevant is that the conjugated operators admit a nice ``mirror" interpretation, visualized in Figure {\bf 5}. The huge Enveloping Algebra induced by the operators entering both $sl(2|1)$ and $sl(2|1)_C$ uniquely determines the spectrum of the theory, as explained in the
comment about Figure {\bf 5}.\par
An enlarged algebraic structure, producing the three-dimensional counterpart of this simpler two-dimensional setting, has to be properly investigated.
\begin{figure}
\begin{centering} 
 \includegraphics[width=0.50\textwidth, angle=0]{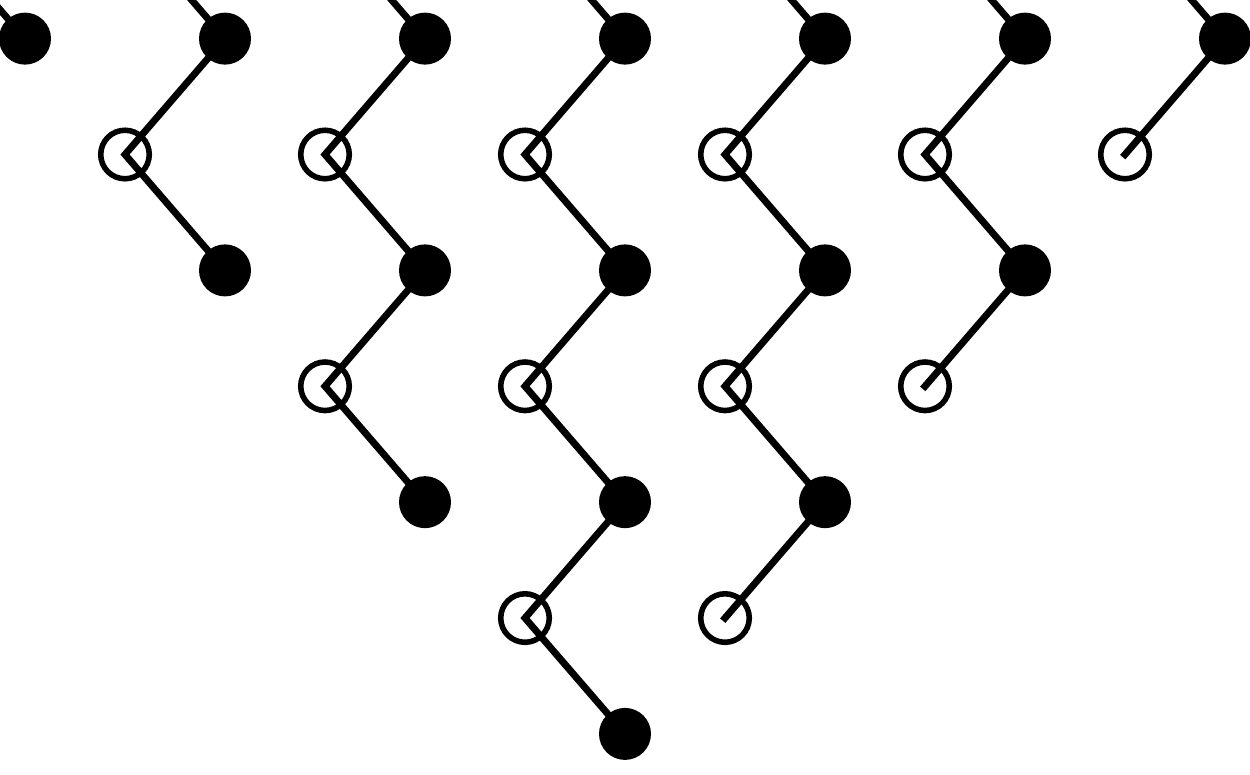}
\begin{center}
{\bf Figure 4:}  Superselected $2D$ oscillator. The bosonic (fermionic) eigenstates are represented by black (white) dots. The $y$ axis labels the integerly spaced energy eigenvalues, while the $x$ axis labels the $so(2)$
spin components. The black dot at the bottom corresponds to the bosonic vacuum. The solid edges represent the action of the creation operator from the $osp(1|2)\subset sl(2|1)$ subalgebra. Infinite $osp(1|2)$ lowest weight representations (a new lowest weight vector at any given energy eigenvalue) are required to produce the spectrum of the theory.$\quad\quad\quad\quad\quad$
\end{center}
~\\
~\\
 \includegraphics[width=0.50\textwidth, angle=0]{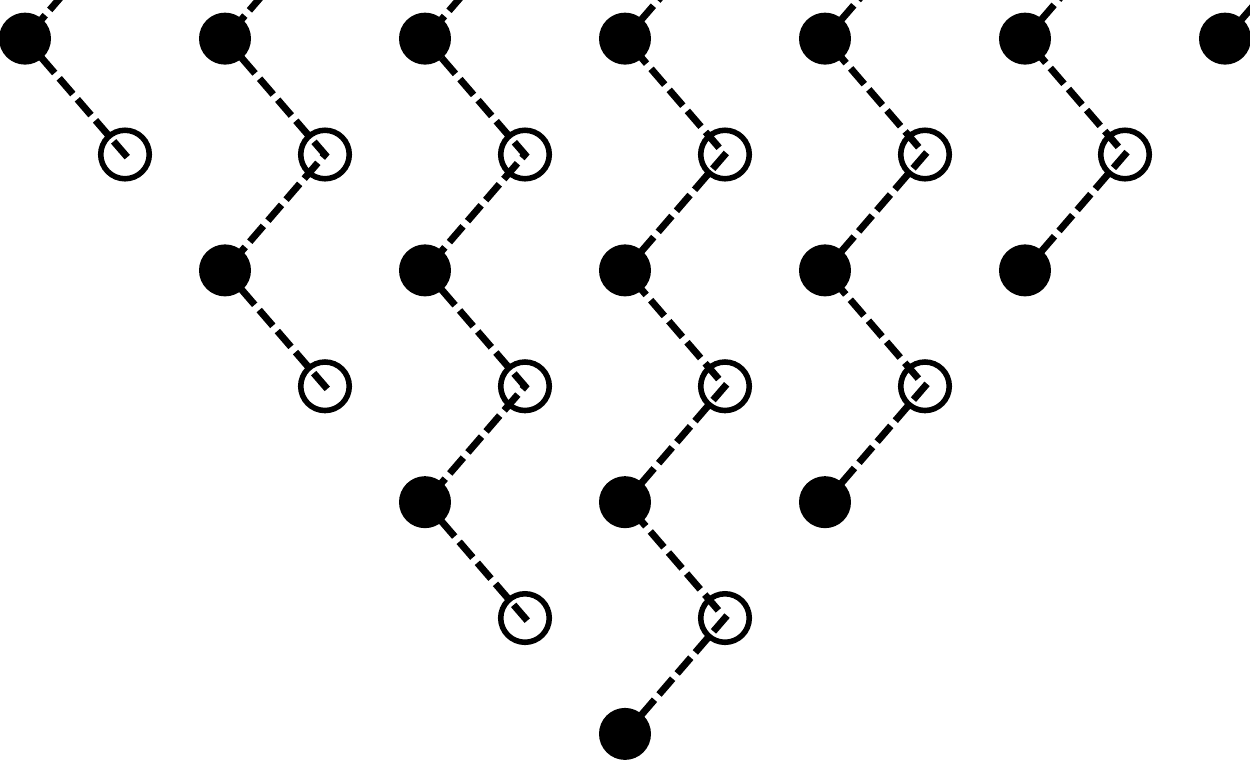}
\begin{center}
{\bf Figure 5:}  A mirror dual of the previous figure. The dashed edges represent the action of the creation operator from  the $osp(1|2)_C\subset sl(2|1)_C$ subalgebra, where $sl(2|1)_C$ is an $sl(2|1)$ superalgebra
produced by a new set of operators obtained by ``mirroring" the operators entering the original $sl(2|1)$ spectrum generating superalgebra.  As before, infinite $osp(1|2)_C$ lowest weight representations are required to produce the spectrum. On the other hand, any energy eigenstate can be obtained from the bosonic vacuum through a path combining both solid and dashed edges.
\end{center}
\end{centering}
\end{figure}
\newpage
\newpage
~\par
~\par
\par {\Large{\bf Acknowledgments}}
{}~\par{}~\par

The authors are grateful to Naruhiko Aizawa and Zhanna Kuznetsova for helpful discussions.  This work is supported by CNPq (PQ grant 308095/2017-0).

\end{document}